\documentclass{aa}

\usepackage{txfonts}
\usepackage{graphicx}
\usepackage{rotating}
\usepackage{natbib}
\usepackage{lscape}

\bibpunct{(}{)}{;}{a}{}{,} 

\newcommand{\ha}{H$\alpha$}
\newcommand{\brg}{Br$\gamma$}
\newcommand{\pab}{Pa$\beta$}
\newcommand{\htwo}{H$_2$}
\newcommand{\hi}{\ion{H}{i}}

\newcommand{\hei}{\ion{He}{i}}
\newcommand{\caii}{\ion{Ca}{ii}}
\newcommand{\oi}{[\ion{O}{i}]}

\newcommand{\av}{$A_V$}

\newcommand{\um}{$\mu$m}

\newcommand{\lsun}{L$_{\odot}$}
\newcommand{\msun}{M$_{\odot}$}
\newcommand{\msunyr}{M$_{\odot}$\,yr$^{-1}$}
\newcommand{\macc}{$\dot{M}_{acc}$}

\newcommand{\lacc}{$L_{\mathrm{acc}}$}
\newcommand{\rstar}{$R_{\mathrm{*}}$}
\newcommand{\ldisk}{$L_{\mathrm{disk}}$}
\newcommand{\lstar}{$L_{\mathrm{*}}$}
\newcommand{\mstar}{$M_{\mathrm{*}}$}
\newcommand{\teff}{$T_\mathrm{eff}$}
\newcommand{\lbol}{$L_{\mathrm{bol}}$}

\begin{document}

\title{POISSON project}

\subtitle{III. Investigating the evolution of the mass accretion rate\thanks{Based on observations collected at the European Southern Observatory, La Silla, Chile (ESO programmes 082.C-0264 and 083.C-0650).}}

\author{S. Antoniucci\inst{1},
		R. Garc\'ia L\'opez\inst{2},	
		B. Nisini\inst{1},
		A. Caratti o Garatti\inst{2},
		T. Giannini\inst{1},
		\and
		D. Lorenzetti\inst{1}
		}

\institute{ INAF-Osservatorio Astronomico di Roma, Via di Frascati 33, 00040 Monte Porzio Catone, Italy \and
			Max-Planck-Institut f\"ur Radioastronomie, Auf dem H\"ugel 69,53121 Bonn, Germany} 

\offprints{Simone Antoniucci, \email{simone.antoniucci@oa-roma.inaf.it}}
\date{Received date / Accepted date}
\titlerunning{Accretion evolution in YSOs}
\authorrunning{Antoniucci et al.}

\abstract
{As part of the POISSON project (Protostellar Optical-Infrared Spectral Survey on NTT), we present 
the results of the analysis of low-resolution near-IR spectroscopic data (0.9-2.4~\um) of two samples 
of young stellar objects in the Lupus (52 objects) and Serpens (17 objects) star-forming clouds, with masses 
in the range of 0.1 to 2.0~\msun\ and ages spanning from 10$^5$ to a few 10$^7$ yr. 
}
{After determining the accretion parameters of the targets by analysing of their \hi\ near-IR emission features,
we added the results from the Lupus and Serpens clouds to those from previous regions (investigated in POISSON with the 
same methodology) to obtain a final catalogue 
(143 objects) of mass accretion rate values (\macc) derived in a homogeneous and consistent fashion.  
Our final goal is to analyse how \macc\ correlates with the stellar mass (\mstar) and how it evolves in time
in the whole POISSON sample.}
{We derived the accretion luminosity (\lacc) and \macc\ for Lupus and Serpens objects from the \brg\ (\pab\ in a few cases) line
by using relevant empirical relationships available in the literature that connect \hi\ line luminosity and
\lacc. To minimise the biases that arise from adopting literature 
data that are based on different evolutionary models and also for self-consistency, we re-derived mass and age for each source 
of the POISSON samples using the same set of evolutionary tracks.}
{We observe a correlation \macc$\sim$\mstar$^{2.2}$ between mass accretion rate and stellar mass, similarly to what 
has previously been observed in several star-forming regions. 
We find that the time variation of \macc\ is roughly consistent with the expected evolution of the accretion rate in 
viscous disks, with an asymptotic decay that behaves as $t^{-1.6}$. However, \macc\ values are characterised by 
a large scatter at similar ages and are on average higher than the predictions of viscous models.}
{Although part of the scattering may be related to systematics due to the employed empirical relationship and to uncertainties
on the single measurements, the general distribution and decay trend of the \macc\ points are real.
These findings might be indicative of a large variation in the initial mass of the disks, 
of fairly different viscous laws among disks, of varying accretion 
regimes, and of other mechanisms that add to the dissipation of the disks, 
such as photo-evaporation.} 

\maketitle

\section{Introduction}
\label{sec:intro}
A significant fraction of the mass of a star is accumulated by accretion from a circumstellar disk.
After a short period of intense accretion, during which the star acquires most of its mass 
(the so-called Class 0 phase), the material continues to be channelled from the inner disk 
onto the central star along the stellar magnetic field lines. Although the rate of mass accretion 
decreases with time, this process plays a crucial role in removing material from the disk along with mass loss, and in this way 
it influences the disk dissipation time-scales and eventually the formation of planets.

The infalling material landing on the stellar surface produces strong shocks and creates a so-called 
hot spot, where the accretion luminosity (\lacc) is radiatively released as continuum and line 
emission. Therefore, by observing this emission one can infer 
quantitative information on the stellar accretion process. 
In particular, \lacc\ can be directly measured for example by modelling the excess of UV continuum emission
shortward of the Balmer and Paschen jumps  (e.g. Calvet \& Gullbring 1998).
Mass accretion rates (\macc) can then be inferred from \lacc\,  if the stellar mass and radius are known, 
in the assumption that the accretion energy is entirely converted into radiation.
It has been found that accretion luminosity derived from the UV continuum excess obeys
empirical relationships with several emission lines such as optical and IR  hydrogen lines, 
CaII, and HeI \citep[e.g.][]{herczeg08, calvet00, calvet04}. The discovery of these correlations
has triggered a significant observational effort to
measure the mass accretion rate of large populations of young stars, with the
aim of defining its dependence on the stellar mass and its time evolution. 

Following first observations concentrated on the population of the Taurus molecular
cloud \citep[e.g.][]{muzerolle98a,herczeg08}, recent studies have been conducted in other star-forming 
regions with the aim of addressing the dependence of mass accretion on the overall age
of the stellar population and environment, such as in $\rho$ Oph \citep{natta04}, L1641 \citep{fang09, caratti12}, 
Chamaeleon, \citep{antoniucci11}, $\sigma$ Ori \citep{rigliaco11a}, and Lupus \citep{alcala14}.
These works have shown that there is a general dependence of \macc\, on both the stellar
mass and age, although with a large scatter. Some of the reasons for this scatter rely 
on accretion variability and uncertainty on the stellar parameters (especially the mass of the objects).
A significant source of scatter also originates from the choice of tracers used to derive \macc\ in the
various studies, however. Indeed, accretion values measured on the same object may differ by more
than one order of magnitude, which can be only partially justified by variability alone
\citep[e.g.][]{costigan12}.
One way to reduce the biases induced by comparing \macc\ values
derived with different instrumentations and in different periods is to perform large
spectroscopic surveys of young stars using (quasi)-simultaneous observations
of tracers at different wavelengths \citep[e.g][]{alcala14}. 

For the POISSON (\textit{Protostellar Objects IR-optical Spectral Survey On NTT}) 
project, we have acquired optical/near-IR low-resolution spectra (R$\sim$700) of 
Spitzer-selected samples of sources in different nearby molecular clouds. In the first paper of the project
\citep[][hereafter Paper I]{antoniucci11} the \lacc\ of sources in the ChaI/II molecular clouds 
was derived from the line luminosity of several optical/IR tracers, and the different 
determinations were then compared. 
In the second paper \citep[][hereafter Paper II]{caratti12}, a sample of young 
sources in the L1641 was characterised 
and their accretion properties were studied, addressing in
particular the dependence of the accretion rate on the age.  

The results so far obtained confirmed that some of the optical tracers commonly adopted
to derive \lacc\ (e.g. H$\alpha$ and Ca II) present
larger discrepancies than other lines, indicating that their
luminosity might be contaminated by contributions in addition to the 
accretion, such as emission from jets and winds. Similar results were recently found by \cite{rigliaco12}.
Among the different tracers, we found that Br$\gamma$ and Pa$\beta$ 
luminosities provide the most reliable value for \lacc in low-mass stars, because they 
require higher excitation conditions than other lines such as \ha, \oi, or \caii\, and they are thus 
less likely to be contaminated by different contributions such as those coming from 
photosphere, winds, and jets. 
In addition, these near-infrared lines are less
dependent on extinction than optical tracers and are particularly suited to derive the
properties of younger and more embedded sources.

In the framework of the POISSON survey, we present here the near-infrared spectra 
(from 0.9 to 2.4$\mu$m) of two samples of young stellar objects (YSOs): one in Lupus (52 objects) and one in Serpens 
(17 sources).
These clouds, which are located at a distance of 150-200 pc and 259 pc
\citep[see][and references therein]{comeron08,straizys96,straizys03},
have been extensively investigated at various wavelengths in recent years:
Hughes et al. (1994), Mer\'in et al. (2008), Comer\'on et al. (2009), Mortier et al. (2011), Alcal\'a et al. (2014) for Lupus,
\nocite{hughes94,merin08,comeron09,mortier11,alcala14} and 
\nocite{winston07,winston09,gorlova10}
Winston et al. (2007, 2009), Gorlova et al. (2010) for Serpens.
These works have provided information about the characterisation and properties of the young 
stellar population of these regions, which have been used
in conjunction with the accretion luminosities and mass accretion rates, mostly derived from the \brg\ 
detected in emission in our spectra.
By combining the Lupus and Serpens YSOs presented here with the previous samples of the POISSON survey, 
we obtain a general catalogue of 143 YSOs with accretion
properties derived in an homogeneous way from near-IR lines.

The paper is structured as follows: in Sect.~\ref{sec:newobs} we present the observations of 
the two samples of Lupus and Serpens and derive the accretion parameters of the sources in
Sect.~\ref{sec:accretion_lup_ser}. Then, in Sect.~\ref{sec:poisson} we 
consider the entire POISSON sample (five star-forming regions) and focus our attention on the temporal
evolution of \macc\ and on its dependence on the stellar mass. Our results are then 
discussed in Sects.~\ref{sec:discuss_relations} and \ref{sec:discuss_macc}.
The main conclusions are summarised in Sect.~\ref{sec:conclusions}.

\begin{table*}[!t]
\caption{Observed Lupus targets and their main properties. All parameters are taken from the literature, except for \mstar, which was recomputed (see text).}
\label{tab:targets_lup}
\begin{tiny}
\begin{tabular}{llcccccccccc}
\hline
\hline
ID  &  Name  &  RA &  DEC  &  Other names  &  \lstar$^a$  &  \teff$^a$  &  \mstar$^a$ & \av &  $L^b_\mathrm{disk}$  &  \lbol & ref. \\ &    &  (2000)  &  (2000)  &    & \lsun  &  K  &  \msun  & mag  &  \lsun  &  \lsun  &  \\ 
\hline
       \multicolumn{12}{c}{\bf{LupI}$^c$} \\ 
\hline
 01  &  Sz65                 &  15:39:27.77  &  -34:46:17.14  &  IK Lup                    &  0.85    &  3800  & 0.52 &  0.2  &  1.01    &  1.86  &2,4\\             
 02  &  Sz66                 &  15:39:28.28  &  -34:46:18.03  &  [MJS2008] 6               &  0.20     &  3415  & 0.31 &  1.0    &  0.94    &  1.14  &1,2\\             
 03  &  [MJS2008] 14         &  15:45:08.87  &  -34:17:33.34  &  ...                       &  0.70     &  4600  & 1.15  &  8.9  &  0.47    &  1.17  &2,5\\             
 04  &  Sz68 A               &  15:45:12.86  &  -34:17:30.60  &  HT Lup                    &  4.82    &  4955  & 2.00 &  1.5  &  0.66    &  5.48  &2\\             
 05  &  Sz69                 &  15:45:17.41  &  -34:18:28.33  &  HW Lup                    &  0.09    &  3197  & 0.19 &  0.    &  0.35    &  0.44  &1,2\\             
 06  &  [MJS2008] 17         &  15:45:18.52  &  -34:21:24.62  &  ...                       &  0.09    &  2700  & 0.06  &  1.6  &  0.01    &  0.10   &2,5\\             
 07  &  Sz71                 &  15:46:44.73  &  -34:30:35.50  &  GW Lup                    &  0.31    &  3632  & 0.42 &  0.5  &  ...       &  ...     &1\\             
 08  &  Sz72                 &  15:47:50.62  &  -35:28:35.34  &  HM Lup                    &  0.25    &  3560  & 0.38 &  0.75 &  ...       &  ...     &1\\             
 09  &  Sz73                 &  15:47:56.94  &  -35:14:34.66  &  HBC600                    &  0.42    &  4060  & 0.80 &  3.5  &  ...       &  ...     &1\\             
 10  &  Sz75                 &  15:49:12.10  &  -35:39:05.12  &  GQ Lup                    &  1.50     &  3900  & 0.59 &  0.95 & ...       &  ...     &4\\             
\hline                                                                                                                                                             
       \multicolumn{12}{c}{\bf{LupII}$^c$} \\                                                                                                                          
\hline                                                                                                                                                             
 11  &  Sz82                 &  15:56:09.22  &  -37:56:05.78  &  IM Lup                    &  1.29    &  3800  & 0.52 &  0.98 &  ...           &  ...     &4\\                 
 12  &  Sz83                 &  15:56:42.30  &  -37:49:15.41  &  RU Lup                    &  1.31    &  4060  & 0.74 &  0    &  ...           &  ...     &1\\              
 13  &  Sz84                 &  15:58:02.53  &  -37:36:02.69  &  ...                       &  0.12    &  3125  & 0.18  &  0    &  ...           &  ...     &1\\               
\hline                                                                                                                                                              
       \multicolumn{12}{c}{\bf{LupIV}$^c$} \\                                                                                                                           
\hline                                                                                                                                                               
 14  &  RY Lup               &  15:59:28.38  &  -40:21:51.30  &  ...                       &  1.26    &  4590  & 1.3 &  0.65 &  ...           &  ...     &4\\                
 15  &  [MJS2008] 146        &  16:00:07.42  &  -41:49:48.42  &  IRAS15567-4141            &  2.74    &  2935  & 0.18  &  2    &  1.93        &  4.67  &3\\              
 16  &  [MJS2008] 149        &  16:00:34.40  &  -42:25:38.62  &  ...                       &  1.45    &  2820  & 0.15  &  2    &  1.18        &  ...     &2\\              
 17  &  [HHC93] F403         &  16:00:44.53  &  -41:55:31.00  &  MY Lup,[MJS2008] 150      &  1.79    &  5152  & 1.39 &  0    &  0.17        &  ...     &2\\              
 18  &  EX Lup               &  16:03:05.49  &  -40:18:25.44  &  HD325367,HBC253           &  0.47    &  3802  & 0.53 &  3.1  &  ...           &  ...     &3,6\\              
 19  &  Sz133                &  16:03:29.39  &  -41:40:01.83  &  ...                       &  0.36    &  4400  & 0.93 &  5.8  &  0.06        &  ...     &5\\             
\hline                                                                                                                                                            
       \multicolumn{12}{c}{\bf{LupIII}$^c$} \\                                                                                                                        
\hline                                                                                                                                                            
 20  &  Sz88 A               &  16:07:00.54  &  -39:02:19.30  &  HO Lup                    &  0.49    &  3850  & 0.57 &  0.25 &  ...           &  ...     &1\\            
 21  &  Sz88 B               &  16:07:00.62  &  -39:02:18.10  &  HO Lup B                  &  0.12    &  3197  & 0.21 &  0    &  ...           &  ...     &1\\            
 22  &  [MJS2008] 20         &  16:07:08.63  &  -39:47:21.90  &  ...                       &  0.30     &  4590  & 0.86 &  1    &  0.08        &  0.38  &2\\             
 23  &  Sz90                 &  16:07:10.07  &  -39:11:03.30  &  [KWS97] Lupus 3 23        &  1.10     &  3900  & 0.60 &  3.2  &  0.46        &  1.56  &2,5\\             
 24  &  Sz95                 &  16:07:52.30  &  -38:58:05.93  &  [MJS2008] 28              &  0.26    &  3400  & 0.31 &  0.7  &  0.07        &  0.33  &2,5\\             
 25  &  [MJS2008] 36         &  16:08:06.18  &  -39:12:22.54  &  ...                       &  2.09    &  2935  & 0.18  &  5.3  &  0.61        &  2.7   &3\\             
 26  &  Sz96                 &  16:08:12.64  &  -39:08:33.48  &  [MJS2008] 37              &  0.82    &  3560  & 0.39 &  1.43 &  0.27        &  1.09  &3\\             
 27  &  Sz97                 &  16:08:21.80  &  -39:04:21.48  &  [MJS2008] 40              &  0.16    &  3270  & 0.25 &  0    &  0.06        &  0.22  &1,3\\             
 28  &  Sz98                 &  16:08:22.49  &  -39:04:46.46  &  HK Lup,V1279 Sco          &  2.35    &  4350  & 1.11 &  2.5  &  1.18        &  3.53  &3\\            
 29  &  Sz99                 &  16:08:24.04  &  -39:05:49.42  &  ...                       &  0.07    &  3270  & 0.21 &  0    &  0.07        &  0.14  &1,3\\             
 30  &  Sz100                &  16:08:25.76  &  -39:06:01.19  &  [MJS2008] 43              &  0.17    &  3057  & 0.17 &  0    &  0.2         &  0.37  &1,3\\             
 31  &  Sz103                &  16:08:30.27  &  -39:06:11.16  &  [MJS2008] 49              &  0.18    &  3270  & 0.25 &  0.7  &  0.08        &  0.26  &1,3\\             
 32  &  [MJS2008] 50         &  16:08:30.70  &  -38:28:26.85  &  ...                       &  3.03    &  5000  & 1.79 &  1    &  0.65        &  3.68  &2\\             
 33  &  Sz104                &  16:08:30.82  &  -39:05:48.87  &  ...                       &  0.10     &  3125  & 0.17 &  0    &  0.06        &  0.16  &1,3\\             
 34  &  HR5999               &  16:08:34.28  &  -39:06:18.16  &  V856 Sco                  &  48.3   &  7890  & 2.55 &  0.85 &  33.99       &  82.28 &2,4\\             
 35  &  Sz106                &  16:08:39.76  &  -39:06:25.32  &  ...                       &  0.10     &  3777  & 0.51 &  1    &  0.11        &  0.21  &1,2\\             
 36  &  Sz107                &  16:08:41.80  &  -39:01:37.02  &  [MJS2008] 58              &  0.15    &  2935  & 0.12 &  0    &  0.01        &  0.16  &3\\             
 37  &  Sz109                &  16:08:48.16  &  -39:04:19.25  &  V1191 Sco                 &  0.14    &  2800  & 0.09  &  0.2  &  0.06        &  0.20   &2,5\\             
 38  &  [CFB2003] Par-Lup3-3 &  16:08:49.40  &  -39:05:39.34  &  ...                       &  0.23    &  3270  & 0.26 &  2.2  &  0.05        &  0.28  &1,2\\             
 39  &  Sz110                &  16:08:51.57  &  -39:03:17.74  &  V1193 Sco,[MJS2008] 67    &  0.28    &  3270  & 0.27  &  0    &  0.07        &  0.35  &1,3\\ 
 40  &  [MJS2008] 68         &  16:08:53.24  &  -39:14:40.17  &  [G2006] 86                &  0.44    &  3900  & 0.62 &  3.3  &  -0.08$^d$       &  0.36  &2,5\\ 
 41  &  Sz111                &  16:08:54.69  &  -39:37:43.11  &  Hen 3-1145,[MJS2008] 71   &  0.33    &  3750  & 0.50 &  0    &  0.2          &  0.53  &1,2\\ 
 42  &  Sz112                &  16:08:55.53  &  -39:02:33.95  &  [MJS2008] 73              &  0.19    &  3125  & 0.20 &  0    &  0.1          &  0.29  &1,2\\ 
 43  &  Sz113                &  16:08:57.80  &  -39:02:22.79  &  [MJS2008] 74              &  0.06    &  3197  & 0.17 &  1    &  0.09         &  0.15  &1,3\\ 
 44  &  Sz114                &  16:09:01.85  &  -39:05:12.42  &  [MJS2008] 80,V908 Sco     &  0.31    &  3175  & 0.23  &  0.3  &  0.4          &  0.71  &1,3\\ 
 45  &  Sz117                &  16:09:44.35  &  -39:13:30.10  &  [MJS2008] 102             &  0.47    &  3700  & 0.46 &  1.5  &  0            &  0.47  &2,5\\ 
 46  &  Sz118                &  16:09:48.65  &  -39:11:16.95  &  [MJS2008] 103             &  0.92    &  4060  & 0.75 &  2.6  &  0.56         &  1.48  &3\\ 
 47  &  [MJS2008] 113        &  16:10:18.58  &  -38:36:12.51  &  ...                       &  0.05    &  2900  & 0.09  &  0.2  &  0.03         &  0.08  &2,5\\ 
 48  &  [MJS2008] 114        &  16:10:19.85  &  -38:36:06.52  &  ...                       &  0.03    &  2990  & 0.09  &  0.4  &  0            &  0.03  &3\\ 
 49  &  Sz123                &  16:10:51.59  &  -38:53:13.77  &  [MJS2008] 121             &  0.20     &  3705  & 0.46  &  1.25 &  -0.01$^d$        &  0.19  &1,2\\ 
 50  &  [MJS2008] 133        &  16:12:11.20  &  -38:32:19.74  &  ...                       &  1.00       &  4400  & 1.20 &  7    &  -0.27$^d$        &  0.73  &2,5\\ 
 51  &  [MJS2008] 136        &  16:12:22.69  &  -37:13:27.65  &  ...                       &  2.87   &  5300  & 1.57 &  6    &  0.36        &  3.23  &2\\ 
 52  &  [MJS2008] 137        &  16:12:43.73  &  -38:15:03.15  &  ...                       &  0.60     &  4400  & 1.10 &  1    &  -0.01$^d$       &  0.59  &2,5\\ 
\hline
\end{tabular}
\end{tiny}
\\~
$^a$ Typical relative uncertainties for the parameters are as follows (see text): \lstar\ 40\%, \teff\ 20\%, \mstar\ 40\%.\\
$^b$ \ldisk\ is computed as \lbol -- \lstar\ (see Sect.~\ref{sec:parameters}).\\
$^c$ Assumed sub-cloud distances are: 150 pc for Lup I, II, and IV and 200 pc for Lup III \citep{comeron08}.\\
$^d$ \ldisk\ is assumed to be 0 in these sources because \lbol--\lstar\ gives a negative value (see text).\\
References. 1: \citet{alcala14}, 2: \citet{merin08}, 3: \citet{mortier11}, 4: \citet{hughes94}, 5: \citet{comeron09}, 6: \citet{lorenzetti12a}.
\end{table*}

\begin{table*}[!t]
\caption{Observed Serpens targets and their main properties. All parameters are taken from the literature, except for \mstar, which was recomputed (see text).}
\label{tab:targets_ser}
\begin{tiny}
\begin{tabular}{llcccccccccc}
\hline
ID  &  Name  &  RA &  DEC  &  Other Names  &  \lstar$^a$   &  \teff$^a$   &  \mstar$^a$   &  \av &  $L^b_\mathrm{disk}$  &  \lbol & ref. \\ 
.  &  .  &  (2000)  &  (2000)  &  .  &  \lsun  &   K  &  \msun  &  mag  &  \lsun  &  \lsun  & \\ 
\hline
\hline
01  &  [WMW2007]103 &  18:29:41.472  &  01:07:37.90  &  ...                 &  1.37  &    4832  & 1.39  & 6.2   &  0.13      &  1.5  &     1,2       \\ 
02  &  [WMW2007]65  &  18:29:43.932  &  01:07:20.80  &  ...                 &  0.21  &    4060  & 0.77  & 8.0   &  0.08      &  0.29 &     1,2       \\ 
03  &  [WMW2007]7   &  18:29:49.602  &  01:17:05.83  &  EC76                &  1.90  &    3400  & 0.32  & 9.6   &  -0.8$^c$  &  1.1  &     1,4       \\ 
04  &  [WMW2007]81  &  18:29:53.608  &  01:17:01.73  &  EC67                &  0.48  &    3560  & 0.38  & 7.4   &  0.31      &  0.79 &     1,2       \\ 
05  &  [WMW2007]38  &  18:29:55.711  &  01:14:31.50  &  GEL4,EC74           &  1.00  &    3199  & 0.24  & 18.5$^d$  &  0.2       &  1.2  &     1,3       \\ 
06  &  [WMW2007]80  &  18:29:56.553  &  01:12:59.65  &  GEL5,SVS4-2,EC79    &  0.49  &    ...   & ...   & 14.2  &  0.1       &  0.59 &     1,3       \\ 
07  &  [WMW2007]85  &  18:29:56.967  &  01:12:47.83  &  EC84                &  2.00  &    2965  & 0.18  & 19.4$^d$  &  ...       &  ...  &     2,3       \\ 
08  &  [WMW2007]35  &  18:29:57.737  &  01:14:05.69  &  GEL10,EC90          &  0.27  &    3270  & 0.26  & 9.6   &  47.73     &  48   &     1,5       \\ 
09  &  [WMW2007]83  &  18:29:57.816  &  01:15:31.86  &  CK13,EC93           &  4.40  &    4900  & 2.00  & 24.4$^d$  &  -1.6$^c$  &  2.8  &     1,2       \\ 
10  &  [WMW2007]70  &  18:29:57.819  &  01:12:28.05  &  [ACA2003]SerA4,EC91 &  1.00  &    3600  & 0.40  & 40.0$^d$  &  0.8       &  1.8  &     1,4       \\ 
11  &  [WMW2007]37  &  18:29:57.849  &  01:12:37.90  &  EC94                &  4.30  &    4000  & 0.70  & 40.0$^d$  &  1.8       &  6.1  &     1,4       \\ 
12  &  [WMW2007]2   &  18:29:57.858  &  01:12:51.40  &  EC92                &  ...   &    ...   & ...   & 9.6   &  ...       &  4.8  &     1         \\ 
13  &  [WMW2007]27  &  18:29:58.205  &  01:15:21.68  &  GEL12,CK4,EC97      &  0.96  &    3997  & 0.69  & 12.8  &  1.64      &  2.6  &     1,2       \\ 
14  &  [WMW2007]4   &  18:29:58.765  &  01:14:25.78  &  EC103               &  ...   &    ...   & ...   & 9.6   &  ...       &  1.6  &     1         \\ 
15  &  [WMW2007]10  &  18:30:02.747  &  01:12:28.06  &  EC129               &  9.50  &    3872  & 0.68  & 24.0$^d$  &  -3.7$^c$  &  5.8  &     1,3       \\ 
16  &  [WMW2007]78  &  18:30:03.413  &  01:16:19.14  &  EC135               &  0.54  &    3850  & 0.57  & 5.9   &  0.2       &  0.74 &     1,2       \\ 
17  &  [WMW2007]73  &  18:30:07.708  &  01:12:04.32  &  ...                 &  0.78  &    3487  & 0.35  & 7.4   &  0.42      &  1.2  &     1,2       \\ 
\hline
\end{tabular}
\end{tiny}
\\~
Assumed distance is 259 pc \citep{straizys96}.\\
$^a$ Typical relative uncertainties for the parameters are as follows (see text): \lstar\ 40\%, \teff\ 20\%, \mstar\ 40\%.\\
$^b$ \ldisk\ is computed as \lbol -- \lstar\ (see Sect.~\ref{sec:parameters}).\\
$^c$ \ldisk\ is assumed to be 0 in these sources because \lbol--\lstar\ gives a negative value.\\
$^d$ For these high-extinction objects the stellar parameters must be considered to be less reliable.\\
References.  1: \citet{evans09}, 2: \citet{winston07, winston09}, 3: \citet{gorlova10}, 4: \citet{doppmann05}, 5: \citet{oliveira09}.
\end{table*}

\section{Lupus and Serpens observations}
\label{sec:newobs}

\subsection{Sample selection}
\label{sec:selection}

Analogously to previous star-forming clouds analysed by POISSON (paper I and II), most of the 
targets in Lupus and Serpens were chosen from the young populations of ClassI/II objects 
identified through Spitzer surveys. 

We used the works by \citet{merin08} and \citet{winston07,winston09} as the main reference articles to select Lupus and Serpens targets.
We selected sources with a $K$-band magnitude $\lesssim$ 12 mag (which was basically imposed by the sensitivity offered by La Silla instruments) and with a spectral energy distribution (SED) spectral index between 2 and 24~\um\footnote{The slope of the spectral energy distribution in this spectral range quantifies the infrared excess and thus provides information on the evolutionary status of the objects. $\alpha_{2-24}$ values were taken from the general catalogue of the \textit{C2D} Spitzer legacy program \citep{evans09}.} $\alpha_{2-24}\gtrsim-1.6$, so as to include the Class II sources as classically defined by \citet{lada84} \citep[see also][]{greene94}.
A few sources with $-1.6<\alpha_{2-24}<-1.9$ or $K$-band magnitude $>$ 12 mag, which were located very close to selected targets, were observed with the same slit acquisition 
and were eventually included in the sample.
By using these selection criteria, which were chosen to also ensure a high detection rate for \hi\ lines, we initially selected 60 objects in Lupus and 30 objects in Serpens, of which we eventually observed 43 in Lupus and 17 in Serpens. 
Finally, we added to the sample nine bright T Tauri objects in Lupus with known disks, taken from the sample of \citet{vankempen07}, which were not present in the list of \citet{merin08}.
The final sample consists of 52 (17) sources in Lupus\footnote{Our Lupus sample is actually composed of objects located in four 
different sub-clouds \citep[Lup I, II, III, and IV, e.g.][see Table \ref{tab:targets_lup}]{comeron08}.} 
(Serpens), namely: 1 (3) Class I, 3 (2) flat, and 48 (12) Class II.

Almost all sources in Lupus are Class II objects, consistent with the relatively 
old average age of 3-4~Myr estimated for the Lupus YSOs \citep{hughes94,mortier11}.
The percentage of younger Class I/flat sources is somewhat higher in the Serpens sample, 
in general agreement with the younger mean age of 1-3 Myr of the Serpens association recently 
found by a few works \citep{winston09,gorlova10}. However, these works also reported 
multiple young stellar populations in the cloud. 

The complete list of targets is presented in Tables \ref{tab:targets_lup} and \ref{tab:targets_ser}.

\subsection{Source parameters}
\label{sec:parameters}

All relevant stellar parameters of the investigated objects (\lstar, \teff, visual 
extinction, disk luminosity \ldisk, and bolometric luminosity \lbol) were taken from the 
literature and are reported in Tables \ref{tab:targets_lup} and \ref{tab:targets_ser} 
together with all the references used.
As a general rule, when more than one literature value 
was available for a given parameter, we adopted the most recent determination, using
the observed dispersion of values to estimate the average uncertainty on that parameter in our sources.

The mean uncertainty on the adopted \lstar\ is a factor 1.4, corresponding 
to 0.1-0.2 \lsun\ for the lowest luminosity objects (\lstar$<$0.7 \lsun).
The typical error on the determination of the spectral type is 1-2 spectral sub-classes, 
which means an uncertainty of a factor 1.1-1.2 for the \teff\ of the sources.

The reported \ldisk\ is an estimate of the luminosity of the circumstellar disk and was simply 
computed as the difference between the total luminosity of the object \lbol\ and the stellar 
luminosity \lstar\ (\ldisk$=$\lbol$-$\lstar).
In a few sources we obtained a negative value of \ldisk, because \lbol\ and \lstar\ values 
are often independent determinations from different works
(inferred by integrating the spectral energy distribution and 
by spectral type fitting). 
In these cases, the derived \ldisk\ is in general a small fraction of \lstar, so we assume that \ldisk\ 
is negligible ($\sim0$~\lsun) in these objects. 

For stellar masses (\mstar), the values derived from the position on the HR diagram 
(\lstar\ and \teff\ values, see Fig. \ref{fig:diag}) were preferred to the ones found in literature.
With this aim, we used the pre-main sequence tracks computed by \citet{siess00}, 
to minimise biases due to the adoption of different pre-main sequence evolutionary models in the reference articles. 
Details about \mstar\ computation are given in Sect.~\ref{sec:global_sample} and in the appendix. 
The relative error on the mass value can be estimated from the uncertainties on \lstar\ and \teff, and is typically 30-40\%.
All sources in both clouds have masses ranging between 0.1 and 2.0 \msun\ and present late 
spectral types, from K to late M, with the exception of HR~5999 in Lupus, which appears as an  
Herbig A6e star of about 2.5\msun.

\subsection{Extinction}
\label{sec:extinction}
Lupus objects present fairly low visual extinction values ($<$\av$>$=1.5 mag), 
while sources in Serpens are on average more embedded, with $<$\av$>=17$ mag.
For Lupus we assumed a mean uncertainty on the extinction of 0.5 mag 
(note that many sources have \av\ values in the range 0-0.5 mag). 
For the more embedded Serpens objects, we considered an \av\ mean uncertainty 
of 1 mag, but assumed a highly conservative uncertainty 
of 4 mag for the six objects with \av$\gtrsim$20 mag.
We considered these uncertainties in computing the accretion parameters (see Sect.~\ref{sec:accretion_lup_ser}).
Because of the large \av\ uncertainty of the high-extinction sources (only four with accretion determinations),
the literature stellar parameters (especially \lstar) must be considered less reliable.
In this work we employed the extinction law parametrisation by \citet{cardelli89}, adopting 
a value of the total-to-selective extinction ratio $R_V$=5.5 for both star-forming clouds.
We used this law to compute (from the provided \av) the extinction at the 
wavelength of the \pab\ and \brg\ lines (i.e. $A_J$ and $A_K$), which we considered to derive the accretion luminosity (see Sect~\ref{sec:lacc}).
Using instead a standard interstellar extinction law like that of \citet{rieke85} to convert $A_V$
would result in a variation of about 20\% of the value of $A_J$ and $A_K$ we obtained from Cardelli's law.

\subsection{Observations and data reduction}
\label{sec:observations}

Observations were carried out at ESO La Silla on 8-11 July 2009 with SofI \citep{sofi,efosc2} mounted on the NTT \citep{ntt}.
We used the red and blue grisms (hereafter RG and BG) of SofI, combined with the 0$\farcs$6 slit, which provides a final resolution R~$\sim$~900 in the 0.9-2.4 \um\, wavelength range (0.94--1.65 \um\ the BG, 1.50--2.40 \um\ the RG). The two grisms were acquired in succession, so that BG and RG spectra can be considered as taken almost simultaneously.
EFOSC2 optical observations of the targets were programmed but could not be performed because of bad weather conditions.

The RG and BG spectra are available for all Lupus sources. As for Serpens, we acquired the RG spectrum for all objects, while the BG was taken only for four sources, which were sufficiently bright ($J\lesssim$14 mag). 

All data were reduced using the IRAF\footnote{IRAF (Image Reduction and Analysis Facility) is a general purpose software system for the reduction and analysis of astronomical data. IRAF is written and supported by the IRAF programming group at the National Optical Astronomy Observatories (NOAO) in Tucson, Arizona. \textit{http://iraf.noao.edu}} software package. 
We followed the standard procedures for bad pixel removal, flat-fielding, and sky subtraction. Spectra of standard stars were acquired at air-masses similar to those of the targets
and used, after removal of any intrinsic line, to correct the scientific spectra for telluric absorption and calibrate the instrumental response. 
Wavelength calibration for all spectra was obtained using xenon-argon arc lamps. 

In addition to spectroscopic data, $H$-band photometry was obtained within 24 hours from the spectral observations for all sources but HR~5999 (see Tables in appendix).
The measured magnitudes were employed to flux-calibrate each RG spectrum at the effective wavelength of the SofI $H$ filter.
2MASS \citep{2mass} $H$ magnitude was instead used to flux-calibrate the RG spectrum of HR~5999. The BG spectrum (when available) was then re-scaled to match the corresponding RG spectrum over the spectral region where the two segments overlap. 

From the flux-calibrated spectra we finally derived $J$- and $K$-band magnitudes by measuring the specific flux at the effective wavelengths of the relative Johnson filters (1.25~\um\ for $J$ and 2.20~\um\ for $K$).
We estimate a typical uncertainty of 0.05 mag for our $H$ photometry measurements and a more conservative uncertainty of 0.1 mag for the $J$- and $K$-band magnitudes, to take into account possible errors on the derived instrument response function. These values translate into a flux calibration relative error of about 10\%, which we assumed for our spectra.
Our photometric results are compared with 2MASS in appendix A for the identification of variable sources.

\subsection{Spectra}
\label{sec:phot_spec}

Some example spectra are shown in Fig.~\ref{fig:spectra} in the Appendix. 
\hi\ recombination lines in emission from the Paschen and Brackett series are the most prominent features detected
in the spectra, in particular the \brg\ and \pab\ lines, which remain unresolved given the low spectral resolution of the observations. 
At least one of these two lines is detected in 30 ($\sim$58\% of the total) sources in Lupus and 8 ($\sim$47\%) in Serpens.
The only other permitted feature observed is the \hei\ line at 1.08\um (19 sources in Lupus), which is a common tracer of winds from the young stars \citep[e.g.][]{edwards06}. Consistently with this origin, we observe an evident P Cygni signature in at least three objects of Lupus.

The detection rates of near-IR \hi\ lines are lower than in clouds investigated in paper I and II with the same instruments and modalities, but are consistent with the Lupus sample being older and Serpens objects being on average more extincted.
Noticeably, as already found for the Cha~I and Cha~II samples of paper I, forbidden emission lines usually taken as indicators of ejection activity \citep[see e.g.][]{nisini05b,podio06} are not detected in any object. Similarly, molecular transitions of \htwo, which is another tracer of shocks induced by outflowing matter, are detected only in one object of each sample.
This lack of jet-line features is probably ascribable more to the low-resolution and relatively low-sensitivity of our observations than to an actual absence of jets from the sources (see Sect. 5 of paper I for a discussion on the sensitivity of the POISSON datasets).

In Tables \ref{tab:lines_lup} and \ref{tab:lines_ser} the fluxes of the \brg\ and \pab\ lines used to derive the accretion luminosity (see Sect.~\ref{sec:lacc}) are given, providing upper limits for non-detections and indicating for each source other emission features at the 3$\sigma$ level.

\section{Accretion in Lupus and Serpens}
\label{sec:accretion_lup_ser}

\begin{figure}[!b] 
\centering
\includegraphics[width=8cm]{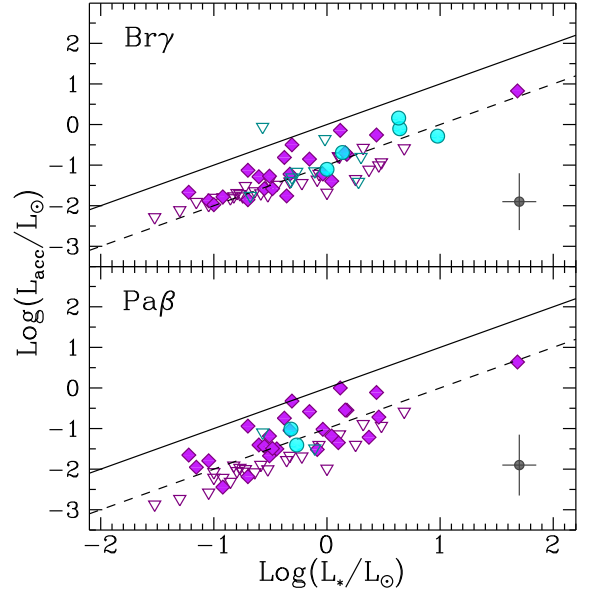}
\caption{\label{fig:lacc_lupser} \lacc\ values derived from \brg\ and \pab\ for Lupus (purple diamonds) and Serpens (cyan circles) sources plotted as a function of \lstar. 
Downward triangles indicate upper limits on \lacc. The solid and dashed lines show the locus where \lacc=\lstar\, and \lacc=0.1\lstar. The mean uncertainty on the single points is also shown. 
}
\end{figure}
\begin{figure}[!b] 
\centering
\includegraphics[width=8cm]{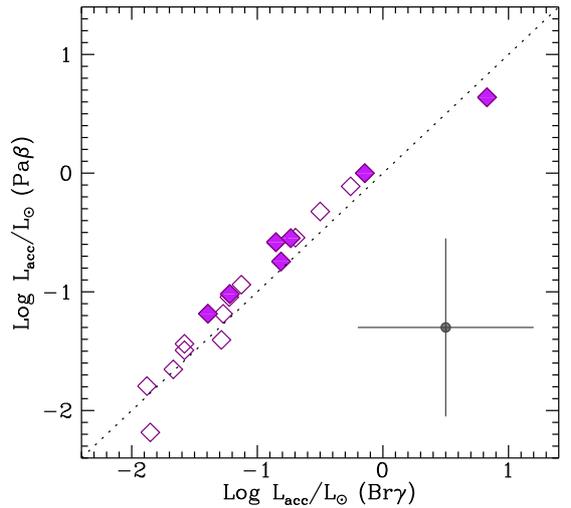}
\caption{\label{fig:lacc_comparison} Comparison between \lacc\ determinations from \brg\ 
and \pab\ for Lupus objects in which both lines are detected. Filled symbols refer to stars with spectral type earlier than M0. 
The dashed line marks the locus of equal accretion luminosity.}
\end{figure}

\begin{table*}[!t]
\caption{\hi\ lines and derived accretion parameters for Lupus targets.}
\label{tab:lines_lup}
\begin{tiny}
\begin{tabular}{l|cc|cc|ccc|l}
\hline
\hline
ID  &  \multicolumn{2}{c|}{\pab}  &  \multicolumn{2}{c|}{\brg} & \lacc(\pab)$^a$ & \lacc(\brg)$^a$ & \macc(\brg)$^a$ & other lines$^b$ \\
  & (F $\pm$ $\Delta$ F)10$^{-14}$ & EW  & (F $\pm$ $\Delta$ F)10$^{-14}$ & EW  &  &  & 10$^{-9}$ &  \\
  & erg s$^{-1}$ cm$^{-2}$ & \AA         & erg s$^{-1}$ cm$^{-2}$ & \AA   & \lsun & \lsun & \msunyr \\
\hline
   01&   $<$  11               &  ...  &      $<$  3.7         &  ...  &        $<$ 0.04  &        $<$ 0.06  &  $<$9.9                         &                         \\
   02&   1.5 $\pm$ 0.4         & -0.7  &  0.7 $\pm$ 0.2    & -0.7  &  0.01            &  0.01            &  2.3                            &                         \\
   03&   5.4 $\pm$ 0.3         & -9.5  &  3.1 $\pm$ 0.3    & -6.3  &  0.25            &  0.14            &  6.3                            &   \hei                  \\
   04&   $<$  47               &  ...  &        $<$  16         &  ...  &        $<$ 0.26  &        $<$ 0.26  &           $<$15                &                         \\
   05&   4.9 $\pm$ 0.6         & -3.5  &  0.7 $\pm$ 0.2    & -1.0  &  0.02            &  0.01            &  2.7                            &                         \\
   06&   $<$  0.54             &  ...  &        $<$  0.4         &  ...   &        $<$ 0.01  &        $<$ 0.01  &           $<$9.5                &                         \\
   07&   5.6 $\pm$ 1.4         & -2.3  &        $<$  2.3         &  ...  &  0.02            &        $<$ 0.04  &                        2.8$^c$  &   \hei                  \\
   08&   9.5 $\pm$ 0.5         & -6.3  &  2.9 $\pm$  0.3    & -4.6  &  0.04            &  0.05            &  7.1                            &   \hei                  \\
   09&   18.0 $\pm$ 1.4        &-12.2  &  6.9 $\pm$  0.9    & -5.2  &  0.18            &  0.15            &  10                            &   \hei                  \\
   10&   61.0 $\pm$ 6.0        & -4.3  &  12.0 $\pm$   3.8    & -1.4  &  0.28            &  0.18            &  34                            &   \hei(PC)              \\
   11&   $<$  16               &  ...  &        $<$  10         &  ...  &        $<$ 0.07  &        $<$ 0.16  &           $<$33                &                         \\
   12&   270.0 $\pm$4.2        &-26.1  &  59.0 $\pm$   2.9    & -9.3  &  1.00            &  0.72            &  90                            &   \hei                  \\
   13&   $<$  1.9              &  ...  &  0.9 $\pm$   0.2    & -2.0  &        $<$ 0.01  &  0.02            &  4.3                            &                         \\
   14&   11.0 $\pm$ 2.7        & -0.7  &        $<$  11         &  ...  &  0.04            &        $<$ 0.17  &                        2.4$^c$  &                         \\
   15&   120.0 $\pm$6.6        & -3.2  &  34 $\pm$   10    & -1.7  &  0.76            &  0.55            &  780                            &                         \\
   16&   45.0 $\pm$ 1.8         & -5.0  &  1.1 $\pm$   1.7    & -2.2  &  0.28            &  0.20            &  270                                   &                         \\
   17&   $<$  12               &  ...  &        $<$  2.8         &  ...  &        $<$ 0.04  &        $<$ 0.05  &  $<$2.2                                   &                         \\
   18&   11.0$\pm$  1.4        & -2.5  &  2.5 $\pm$  0.6    & -2.0  &  0.09            &  0.06            &  7.1                            &   \hei                  \\
   19&   1.7 $\pm$  0.2        & -2.8  &        $<$  0.9         &  ...  &  0.03            &        $<$ 0.03  &                        1.4$^c$  &   \hei                  \\
   20&   69.0 $\pm$ 1.7        &-20.3  &  13.0 $\pm$   0.7    & -8.6  &  0.47            &  0.32            &  35                            &   \hei                  \\
   21&   0.65 $\pm$ 0.2        & -0.7  &        $<$  0.5         &  ...  &  0.00            &        $<$ 0.02  &                        0.8$^c$  &                         \\
   22&   $<$  1.3               &  ...  &        $<$  0.5         &  ...   &        $<$ 0.01  &        $<$ 0.02  &           $<$0.7                &                         \\
   23&   4.2 $\pm$  0.7        & -2.2  &  0.9 $\pm$   0.2    & -0.8  &  0.06            &  0.04            &  6.2                            &   \hei(PC)              \\
   24&   $<$  1.8               & ...   &        $<$  0.6         &  ...  &        $<$ 0.01  &        $<$ 0.02  &           $<$3.8                &                         \\
   25&   $<$  4.4              & ...  &        $<$  5.7         &  ...  &        $<$ 0.13  &        $<$ 0.26  &           $<$330                &                         \\
   26&   3.4 $\pm$  0.8        & -1.5  &        $<$  1.2         &  ...  &  0.03            &        $<$ 0.04  &                        7.5$^c$  &                         \\
   27&   $<$  1.6               & ...   &        $<$  0.6         &  ...   &        $<$ 0.01  &        $<$ 0.02  &           $<$4.2                &                         \\
   28&   4.9 $\pm$  0.6        & -3.0  &        $<$  2.0         & ...   &  0.06            &        $<$ 0.08  &                        5.9$^c$  &   \hei                  \\
   29&   1.9 $\pm$  0.3        & -3.3  &        $<$  0.4         &  ...   &  0.01            &        $<$ 0.01  &                        1.8$^c$  &                         \\
   30&   $<$  1.9               &  ...  &        $<$  0.6         & ...   &        $<$ 0.01  &        $<$ 0.02  &           $<$6.7                &                         \\
   31&   $<$  1.2               &  ...   &        $<$  0.5         &  ...  &        $<$ 0.01  &        $<$ 0.02  &           $<$3.6               &                         \\
   32&   $<$  14               & ...   &        $<$  3.9         &  ...  &        $<$ 0.12  &        $<$ 0.12  &           $<$6.1                &                         \\
   33&   $<$  1.1               &  ...  &        $<$  0.5         &   ... &        $<$ 0.01  &        $<$ 0.02  &           $<$4.0                &                         \\
   34&   500 $\pm$  78         & -4.5  &  45 $\pm$  10   & -0.6  & 4.33             &   6.73$^d$ &  380$^d$     &   \hei                  \\
   35&   $<$  1.1               &  ...  &  0.3 $\pm$  0.1    & -0.7  &        $<$ 0.01  &  0.01            &  0.6                            &   \hei, \htwo           \\
   36&   $<$  2.1               &  ...  &        $<$  0.5         &  ...  &        $<$ 0.01  &        $<$ 0.02  &           $<$8.2                &                         \\
   37&   $<$ 0.81               &  ...   &       $<$ 0.4         &   ... &        $<$ 0.01  &        $<$ 0.02  &           $<$4.4                &                         \\
   38&   $<$  0.77              &  ...  &        $<$ 0.5         &   ... &        $<$ 0.01  &        $<$ 0.02  &           $<$5.2                &                         \\
   39&   6.2 $\pm$  0.6         & -4.6  &  0.8 $\pm$   0.1    & -1.8  &  0.04            &  0.03            &  4.8                            &   \hei                  \\
   40&   $<$  1.1               &   ... &  0.3 $\pm$   0.1    & -0.8  &        $<$ 0.02  &  0.02            &  1.6                            &                         \\
   41&   5.5 $\pm$  0.6         & -2.9  &  0.8 $\pm$   0.2    & -1.4  &  0.03            &  0.03            &  2.9                            &   \hei                  \\
   42&   $<$  1.8               &  ...  &        $<$  1.0         &  ...  &        $<$ 0.01  &        $<$ 0.03  &           $<$9.2                &                         \\
   43&   2.8 $\pm$  0.2         & -9.1  &  0.6 $\pm$  0.1    & -4.2  &  0.02            &  0.02            &  3.9                            &   \hei                  \\
   44&   9.9 $\pm$  1.0         & -4.5  &  1.8 $\pm$  0.4   & -2.3  &  0.07            &  0.05            &  17                            &   \hei                  \\
   45&   $<$  2.4               &  ...  &        $<$  1.1         &  ...  &        $<$ 0.02  &        $<$ 0.04  &           $<$5.6                &                         \\
   46&   7.3 $\pm$  0.7         & -3.5  &  1.5 $\pm$  0.5    & -1.2  &  0.09            &  0.06            &  6.2                            &                         \\
   47&   $<$  0.32               &  ...  &        $<$ 0.2         &  ...  &        $<$ 0.01  &        $<$ 0.01  &           $<$3.1                &   \hei(PC)              \\
   48&   $<$  0.22               &  ...  &        $<$ 0.1         &  ...  &        $<$ 0.01  &        $<$ 0.01  &           $<$1.5                &                         \\
   49&   13.0 $\pm$ 0.5         &-10.9  &  2.3 $\pm$  0.2    & -4.6  &  0.11            &  0.07            &  7.0                            &   \hei                  \\
   50&   $<$  0.23               &  ...  &        $<$ 0.2         &  ...  &        $<$ 0.01  &        $<$ 0.02  &           $<$1.2                &                         \\
   51&   5.3 $\pm$  0.4         & -2.7  &        $<$  1.7         &  ...  &  0.18            &        $<$ 0.10  &                        9.4$^c$  &                         \\
   52&   $<$  2.6               &  ...  &        $<$  1.1         &  ...  &        $<$ 0.02  &        $<$ 0.04  &           $<$1.8               &                         \\
\hline
\end{tabular}
\end{tiny}
~\\
 $^a$ Typical uncertainties on \lacc\ and \macc\ is 0.7 dex and 0.8 dex.\\
 $^b$ P Cygni profiles are indicated with PC.\\
 $^c$ \macc\ value computed from \pab.\\ 
 $^d$ Value obtained considering correction for \brg\ photospheric absorption (see text).\\
\end{table*}

\begin{table*}[!t]
\caption{\hi\ lines and derived accretion parameters for Serpens targets.}
\label{tab:lines_ser}
\begin{tiny}
\begin{tabular}{l|cc|cc|ccc|l}
\hline
\hline
ID  &  \multicolumn{2}{c|}{\pab}  &  \multicolumn{2}{c|}{\brg} & \lacc(\pab)$^a$ & \lacc(\brg)$^a$ & \macc(\brg)$^a$ & other lines \\
  & (F $\pm$ $\Delta$ F)10$^{-14}$ & EW  & (F $\pm$ $\Delta$ F)10$^{-14}$ & EW  &  &  & 10$^{-9}$ &  \\
  & erg s$^{-1}$ cm$^{-2}$ & \AA         & erg s$^{-1}$ cm$^{-2}$ & \AA   & \lsun & \lsun & \msunyr \\
\hline
01  &           ...                     &  ...     &   2.2  $\pm$  0.1       &  -3.5      &    ...      &    0.20                        &  9.7                &           \\
02  &           ...                     &  ...     &                     $<$  0.1  &   ...      &    ...      &              $<$ 0.02          &  $<$0.8             &           \\
03  &           ...                     &  ...     &                     $<$  0.2  &   ...      &    ...      &              $<$ 0.04          &  $<$19             &           \\
04  &         1.1 $\pm$ 0.2   & -4.0     &                     $<$  0.3  &   ...      &   0.09      &              $<$ 0.04          &  18$^{b}$          &           \\
05  &           ...                     &  ...     &   0.20  $\pm$  0.06       &  -1.3      &    ...      &    0.08                        &  40                &           \\
06  &           ...                     &  ...     &                     $<$  0.2  &   ...      &    ...      &              $<$ 0.05          &          ...               &           \\
07  &           ...                     &  ...     &                     $<$  0.3  &   ...      &    ...      &              $<$  0.2          &  $<$180             &           \\
08  &          $<$ 0.5          &  ...     &                     $<$  7.3  &   ...      &   $<$0.08   &              $<$  0.8          &  $<$210             &           \\
09  &           ...                     &  ...     &  1.0  $\pm$  0.3       &  -6.8      &    ...      &     0.74                       &  43                &           \\
10  &           ...                     &  ...     &                     $<$  8.4  &   ...      &    ...      &              $<$ 0.01          &  $<$0.03             &           \\
11  &           ...                     &  ...     &  0.30  $\pm$ 0.06       &  -2.6      &    ...      &      1.32                      &  320                &           \\
12  &           ...                     &  ...     &                     $<$  0.4  &   ...      &    ...      &              $<$ 0.06          &       ...                  &           \\
13  &           ...                     &  ...     &                     $<$  2.3  &   ...      &    ...      &              $<$ 0.43          &  $<$51             &           \\
14  &           ...                     &  ...     &   0.20  $\pm$  0.02       &  -5.0      &    ...      &    0.03                       &       ...                  &  \htwo    \\
15  &           ...                     &  ...     &   0.7  $\pm$   0.06       &  -1.8      &    ...      &     0.49                       &  200                &           \\
16  &         0.7 $\pm$ 0.2   & -1.8    &                     $<$  0.7  &   ...      &   0.04      &              $<$ 0.07          &  4.4$^{b}$          &           \\
17  &          $<$ 0.4          &  ...  &                     $<$0.7  &   ...      &   $<$0.03   &              $<$ 0.07          &  $<$19             &           \\
\hline
\end{tabular}
\end{tiny}
~\\
$^a$ Typical uncertainties on \lacc\ and \macc\ is 0.7 dex and 0.8 dex.\\
$^b$ \macc\ value computed from \pab.
\end{table*}

\subsection{Accretion luminosities}
\label{sec:lupser_lacc}

Consistently with previous POISSON papers, we derived the accretion luminosity (\lacc) from \hi\ lines 
using the empirical relationships provided by \citet{calvet00,calvet04}, which connect the \brg\ (or \pab) luminosities to \lacc:
\begin{equation}
\label{eq:calv1}
\mathrm{Log}\;L_{acc}/L_\odot = 0.9 \cdot \mathrm{Log}\;L_{\mathrm{Br\gamma}}/L_\odot + 2.9 ~,
\end{equation}
\begin{equation}
\label{eq:calv2}
\mathrm{Log}\;L_{acc}/L_\odot = 1.03 \cdot \mathrm{Log}\;L_{\mathrm{Pa\beta}}/L_\odot + 2.8 ~. 
\end{equation}
The \hi\ line luminosities were computed from measured fluxes corrected for extinction. 
The \lacc\ values inferred from the above relationships are listed in Tables~\ref{tab:lines_lup} 
and \ref{tab:lines_ser}, where we also provide upper limits (3$\sigma$) for non-detections.
We discuss in Sect.~\ref{sec:discuss_relations} how these results vary by adopting  
other empirical relationships available in the literature.

In Fig.~\ref{fig:lacc_lupser} we plot \lacc\ determinations from \brg\ and \pab\ as 
a function of the stellar luminosity \lstar.
As for previously investigated samples (paper I and II), the plots show that \lacc\ is correlated 
with \lstar\ and that the accretion luminosity is only a fraction of the stellar luminosity
(as expected for Class II objects), with 0.1~\lstar $<$ \lacc $<$ \lstar\ for most sources,
except for a few low accretors (0.01~\lstar $<$ \lacc $<$ 0.1~\lstar) found in particular in Lupus.

For consistency with the procedure employed in paper I (see discussion in Sect.~6 therein), 
in this work we adopt \lacc\ values (and upper limits) computed from the luminosity of \brg. 
We remark that we did not take into account the stellar photospheric absorption for
the lines, whose contribution to \brg\ is negligible for late spectral types.
The only exception is the source HR5999 in Lupus (spectral type A6), for which we computed the flux 
correction on the basis of the \brg\ absorption observed in a template spectrum of the same spectral type
\citep[for details see][]{garcia_lopez11}.
The accretion luminosities were derived from \pab\ only for those 
sources where \brg\ is not detected (eight objects in Lupus and two in Serpens, see Tables~\ref{tab:lines_lup} and \ref{tab:lines_ser}).
Neglecting photospheric absorption of \pab\ for stars earlier than M0 might lead to underestimate line fluxes (and hence accretion rates),
this affects four of the eight sources of Lupus with \lacc\ derived from \pab.
For sources where both lines are observed (only 19 sources in Lupus)
a direct comparison between the two \lacc\ from \brg\ and \pab\ 
shows that the values agree well within uncertainties on \lacc\ values, also in objects with spectral types earlier than M0, 
as displayed in Fig.~\ref{fig:lacc_comparison}.
At variance with Cha~I and Cha~II results (see Fig.~4 in paper I), in Lupus objects 
we therefore find no evidence of sources for which \pab\ provides \lacc\ values significantly 
underestimated with respect to \brg. Accordingly, measured intrinsic \pab/\brg\ 
ratios are similar to those observed in Taurus (\pab/\brg~$\gtrsim3$, Muzerolle et al. 1998). 
\nocite{muzerolle98a}

The error on the derived \lacc\ values depends on the errors on both line flux and extinction, but it is largely dominated by the uncertainty 
on the parameters of the empirical relationships employed\footnote{This latter reflects the intrinsic scatter of the points over which the relationship was fit, which can be estimated to be about 0.65 dex in the range of line luminosities of our objects \citep[see][]{muzerolle98a,calvet00,calvet04}.}, so that we estimate it to be about 0.7 dex.

\begin{figure*}[t] 
\centering
\includegraphics[width=3.8cm]{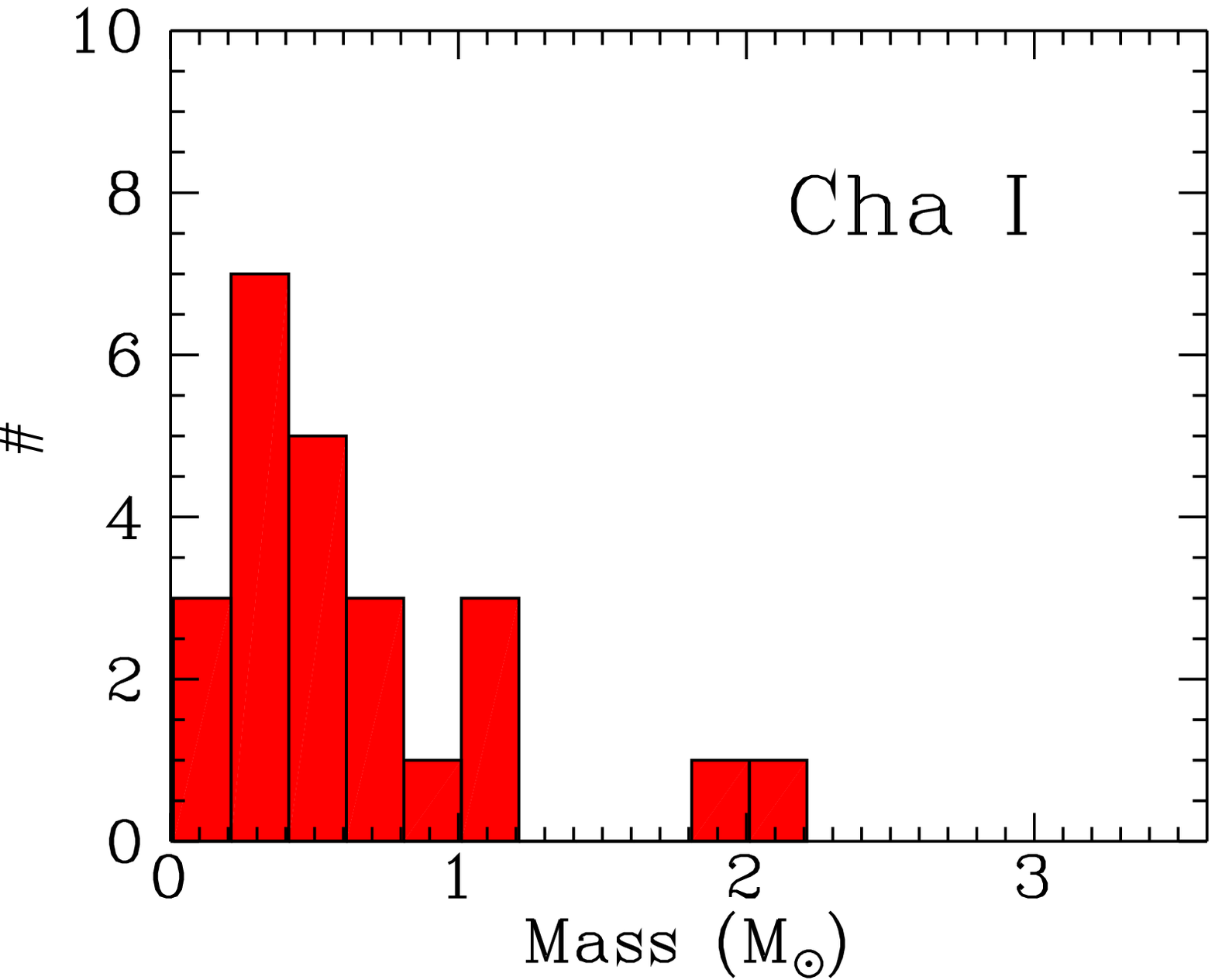}\hspace{-.3cm}
\includegraphics[width=3.8cm]{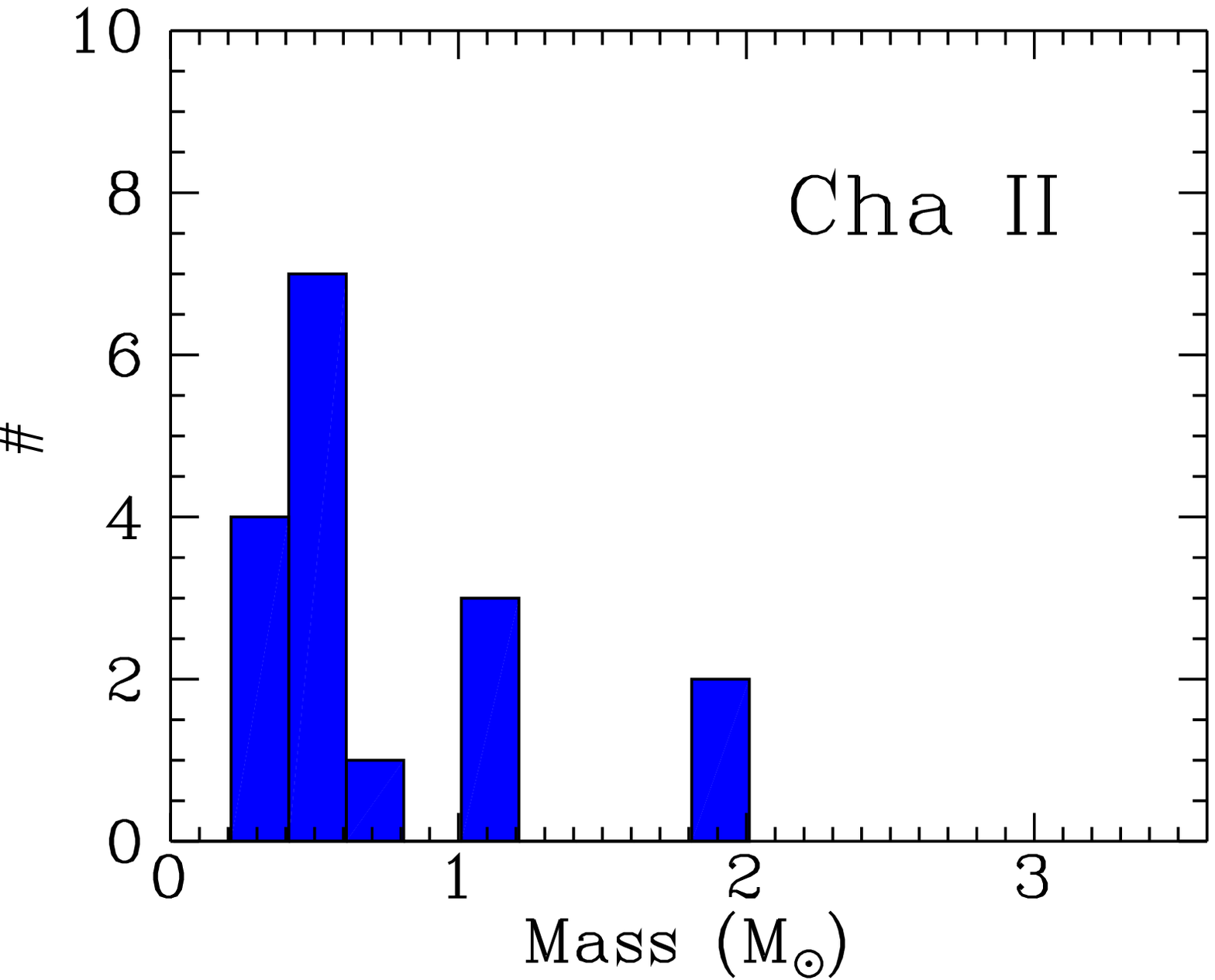}\hspace{-.3cm}
\includegraphics[width=3.8cm]{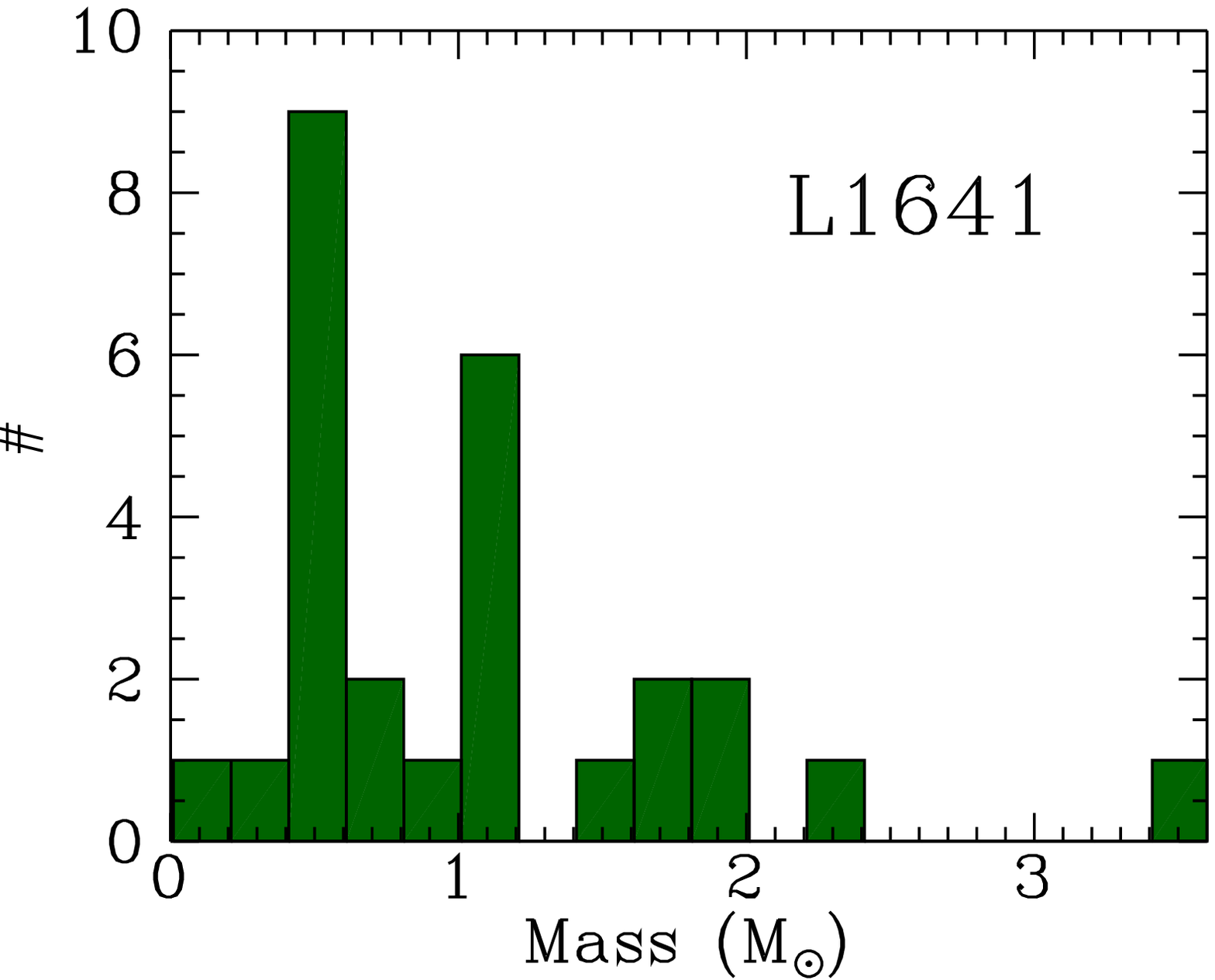}\hspace{-.3cm}
\includegraphics[width=3.8cm]{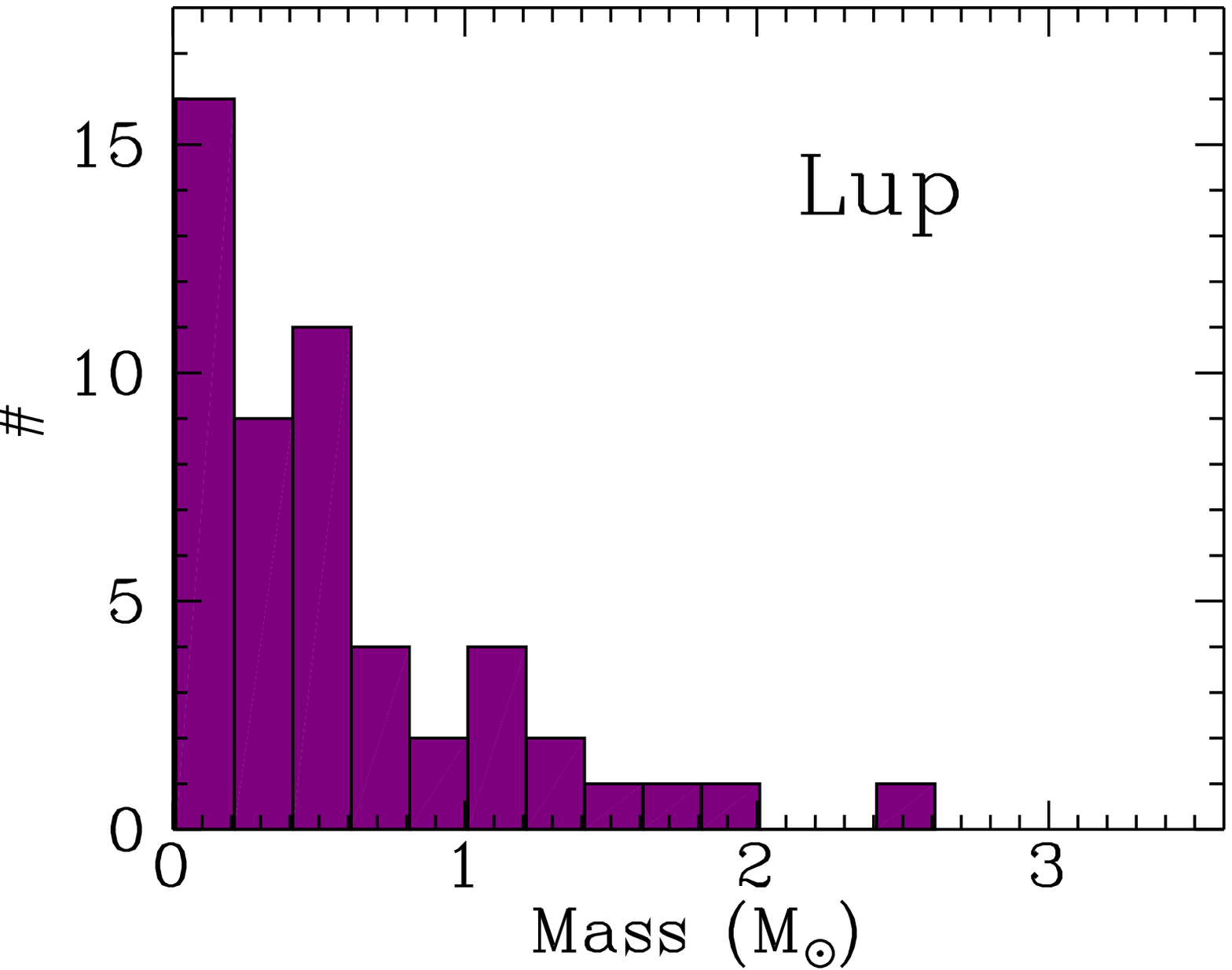}\hspace{-.3cm}
\includegraphics[width=3.8cm]{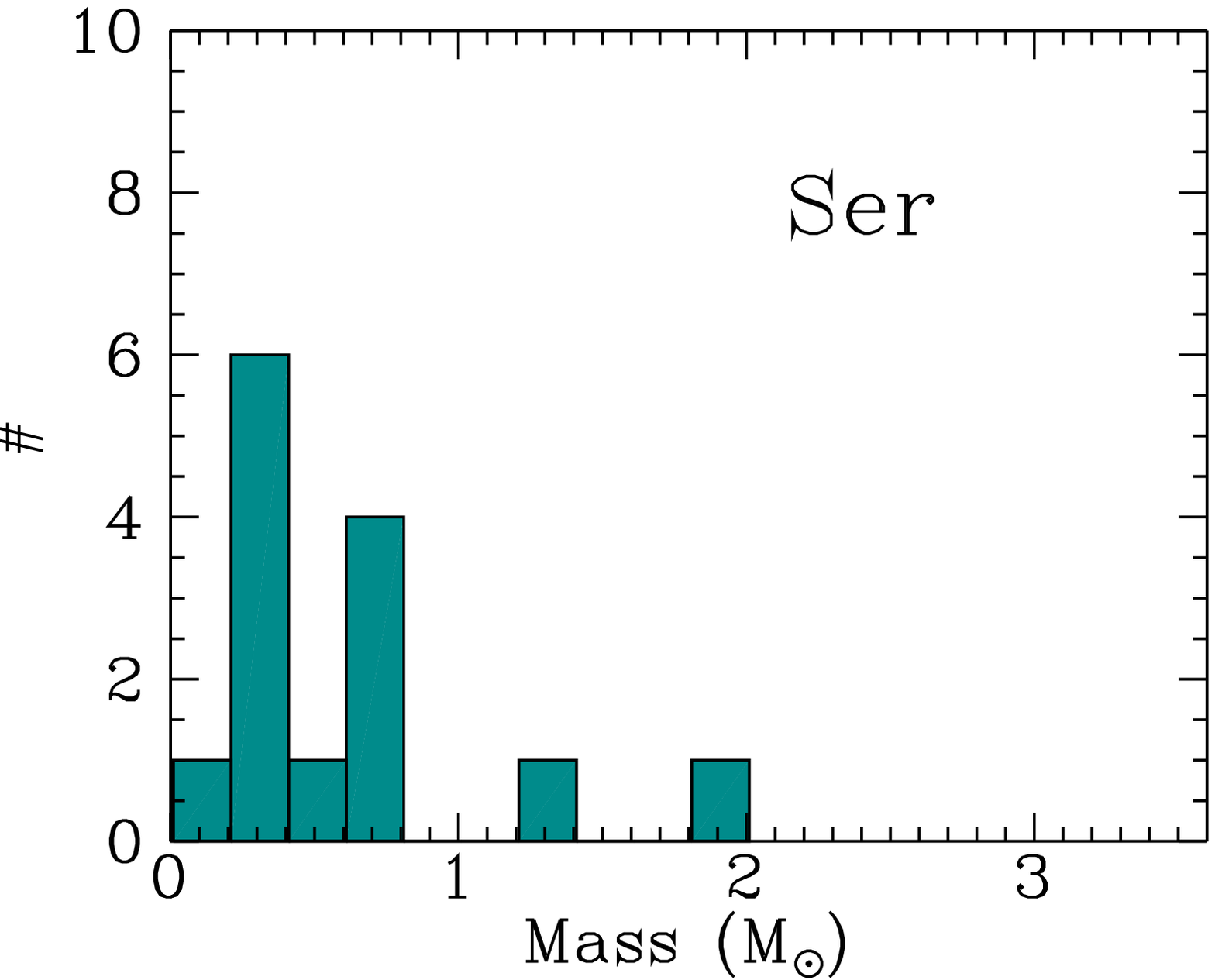}
\caption{\label{fig:mass_histo} \textbf{Histograms of stellar masses for the five samples investigated by POISSON}.
}
\end{figure*}

\subsection{Mass accretion rates}
\label{sec:lupser_macc}

From the accretion luminosity \lacc\ we computed the mass accretion rate (\macc) by using 
the relationship \citep[e.g.][]{gullbring98}
\begin{equation}
\label{eq:macc}
\dot{M}_{acc} = \frac{L_{acc}R_{*}}{GM_{*}} \left(1-\frac{R_{*}}{R_{in}}\right)^{-1} ~,
\end{equation}
where \mstar\ and \rstar\ are the stellar mass and radius, and $R_{in}$ is the inner 
truncation radius of the disk, which we assumed equal to $5 R_{*}$.
Mass estimates were derived as explained in Sect.~\ref{sec:global_sample} and the appendix, while
\rstar\ was calculated from \lstar\ and \teff\ listed in Tables \ref{tab:targets_lup} and \ref{tab:targets_ser}:
\begin{equation}
R_{*}=\sqrt{L_{*}/4\pi \sigma T_\mathrm{eff}^4} ~.
\end{equation}
The \macc\ values thus obtained are reported in Tables \ref{tab:lines_lup} and \ref{tab:lines_ser} 
for Lupus and Serpens. These are about 10$^{-8}$-10$^{-9}$ \msunyr for all objects, except for five sources 
displaying \macc $\sim$10$^{-7}$ \msunyr (three in Lupus and two in Serpens).
We remark that two of the objects in Lupus are HR5999 (a Herbig star for which a higher accretion rate is expected) and [MJS2008]~146, 
which conversely is a low-mass source (0.18 \msun) that appears to be in an enhanced phase of accretion (see below).
The inferred rates are in the range of values commonly observed in many Class II objects 
\citep[see e.g. paper I, ][]{gullbring98,natta06,white07}.
Considering the uncertainties on \lacc\ and \mstar, and assuming 
an additional typical 30\% error on \rstar, we derive a relative error on the accretion rates 
of about 0.8 dex.

Mass accretion rates for several Lupus sources of our sample were recently inferred by \citet{mortier11} 
from the analysis of VLT/FLAMES spectra and by Alcal\'a et al. (2014)\nocite{alcala14}
from VLT/X-Shooter data.
Alcal\'a et al. (2014) observed 22 objects of our Lupus sample and
derived \macc\ estimates on the basis of the detected Balmer jump. 
The authors also measured the flux of several emission lines in the optical-NIR range 
(including \brg\ and \pab) and used these results to provide new calibrated \lacc-line relationships.
From examining their \brg\ fluxes, we find a reasonable agreement with our measurements because fluxes differ by
a factor lower than 3 in almost all objects.
Albeit the fluxes are similar, the \macc\ we eventually obtain appears to be on average higher than those 
derived in \citet{alcala14} because of the \lacc-\brg\ flux relationship we employed (see discussion in Sect.~\ref{sec:discuss_relations}).

\citet{mortier11} give \macc\ based on \ha\ flux and width for 14 sources of our Lupus sample and 
seven for which we have a \brg\ detection. Our \macc\ are very similar only for three of these 
sources, while they are significantly different for the others, with two targets showing a discrepancy 
exceeding 2 orders of magnitude. 

Analogously, in eight sources in common between Mortier et al. and Alcal\'a et al., we 
find that in three cases their \macc\ determinations differ by more than one order of 
magnitude. These fluctuations are similar to those observed in paper I 
for \lacc\ values derived from \ha, which we judged as the less reliable tracer.
Indeed, \lacc\ determinations based on \ha\ were characterised by a 
scatter two orders of magnitude greater than those from \brg, most likely due to different emission components contributing
to the line flux (see paper I).

For [MJS2008]~146, we measure an accretion rate about four 
orders of magnitude higher than Mortier et al., although using a different tracer. 
This enormous variation suggests that this object undergoes phases of outburst 
and quiescence, and as such it appears to be a good EXor candidate \citep[e.g.][]{lorenzetti12a,antoniucci13d}.
This interpretation is also supported by the significant variation in magnitude with respect 
to 2MASS ($\Delta J \sim 1$ mag), which we signalled in Appendix A. 

For Serpens sources no previous \macc\ determinations were available in the 
literature.

\begin{table}[!t]
\caption{\label{tab:numbers} Number of observed objects compared with the total number of YSOs in the five samples of POISSON, with upper and lower limits of \mstar.}
\begin{tiny}
\begin{tabular}{lccc}
\hline
Region        &  total \# of Class I/II YSOs$^{a}$  &  observed  & \mstar\ range (\msun)     \\
\hline
\hline
Chamaeleon I  &  108          & 30                &  0.2--2.0            \\
Chamaeleon II &  26           & 17                &  0.2--1.9            \\
L1641         &  $\sim$200    & 27                &  0.2--3.5            \\
Lupus         &  95           & 52                &  0.1--2.0            \\
Serpens       &  117          & 17                &  0.2--2.6            \\
\hline
\end{tabular}
\end{tiny}
\\~
$^{a}$ Estimates of the number of YSOs in the clouds are taken from \citet[][ChaI]{luhman07}, \citet[][ChaII]{alcala08}, Allen \& Davis (2008, L1641), \citet[][Lup, Ser]{evans09}.
\end{table}

\section{Accretion parameters of the whole POISSON sample}
\label{sec:poisson}

\subsection{Whole sample}
\label{sec:global_sample}

The analysis of the POISSON spectra (this work and papers I and II) provides us with
accretion luminosities and rates for a total of 143 young objects located in five different star-forming regions.
In Table~\ref{tab:numbers} we report the total number of objects observed per cloud compared with the estimated total number of Class I, Flat, and Class II YSOs
in each region, as well as the interval of masses explored. 
The histograms of the stellar masses for each sample is shown in Fig.~\ref{fig:mass_histo}. 
Because of the somewhat stringent selection criteria we adopted, POISSON was in general poorly sensitive to stars with 
masses below $\sim$0.1-0.2\msun. The mass distribution is similar for the Serpens and Chamaeleon clouds, while Lupus, which is the richest sample and has
the lowest extinction values, presents more low-mass objects. 
Converselyy, the L1641 sample has a smaller fraction of low-mass objects than other regions, because of 
its greater distance (the selection limit of $K \lesssim 12$ was adopted for all clouds) and its higher mean extinction.
Most of the POISSON objects have masses in the range of 0.2 to 1.5~\msun\ 
and age estimates spanning from about 10$^4$ to a few 10$^7$ yr.

In the following sections, we relate the accretion parameters derived in a homogeneous fashion across the whole sample
to source properties such as disk and stellar luminosity, mass, and age.
To this aim, it is therefore important to ensure that the source parameters have been derived 
in a fashion as consistent as possible for all the sub-samples.
This is certainly true in our large sample, where the \lacc\ values have been computed using the same tracer (i.e. \hi\ \brg) 
and relationship \citep{calvet04}\footnote{For self-consistency, the \lacc\ values provided here for the L1641 objects are  
derived from \brg\ alone, while the numbers reported in paper II are an average of \lacc(\pab) and \lacc(\brg) values.}.
As already mentioned, \lacc\ from the \pab\ line was adopted 
only for those few sources where \brg\ was not detected (see Sect.~\ref{sec:accretion_lup_ser}).

As for accretion rates, \macc\ values depend also on the stellar mass of the objects 
(see Eq.~\ref{eq:macc}), which is usually derived from the position 
in the HR diagram and from a comparison with predictions of evolutionary models. 
Because it is well known that \mstar\ may significantly vary depending on the assumed model 
(see e.g. discussion in Spezzi et al. 2008\nocite{spezzi08}), we decided to re-derive the stellar masses adopting 
the set of evolutionary models of \citet{siess00} for all the POISSON objects, 
to minimise the biases on \macc\ determinations.
An important by-product of this procedure is a homogeneous and consistent set of age estimates, 
which are inferred from comparison with the isochrones of the models developed by Siess et al..
Considering the average uncertainties (for the positioning over the HR diagram), we may assume that these age
estimates are correct within a factor 2.
More details about this procedure are given in the appendix, where the HR diagram for the 
whole sample is reported (Fig.~\ref{fig:diag}),
along with all the relevant parameters 
(\lstar, \teff, \mstar, age, and \macc) of the POISSON targets (Tables \ref{tab:hr1} and \ref{tab:hr2}).

\subsection{Accretion luminosities}
\label{sec:lacc}

\begin{figure}[!t] 
\centering
\includegraphics[width=8cm]{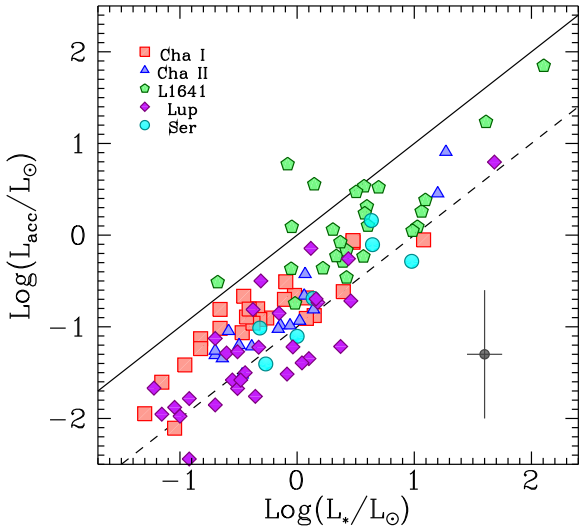}
\caption{\label{fig:lacc_lstar_all} Accretion luminosity as a function of the stellar luminosity 
for the whole POISSON sample. The various symbols refer to the five different cloud samples (see legend).
The solid and dashed lines show the locus of \lacc=\lstar\, and \lacc=0.1\lstar.
}
\end{figure}

Derived accretion luminosities for the POISSON sources are depicted as a function of \lstar\ 
in Fig.~\ref{fig:lacc_lstar_all}, where it is evident that a correlation 
between \lacc\ and \lstar\ is found in all sub-samples. 
Most sources show \lacc\ values in the range (0.1)~\lstar\ and 1~\lstar, except for some low-accretors (mostly in Lupus) and a few L1641 objects 
that conversely show enhanced accretion activity. One of these latter is the source
V2775 Ori (CTF 216-2), which was in an outbursting phase during the observations \citep{caratti11,caratti12}.

\begin{figure}[] 
\centering
\includegraphics[width=8cm]{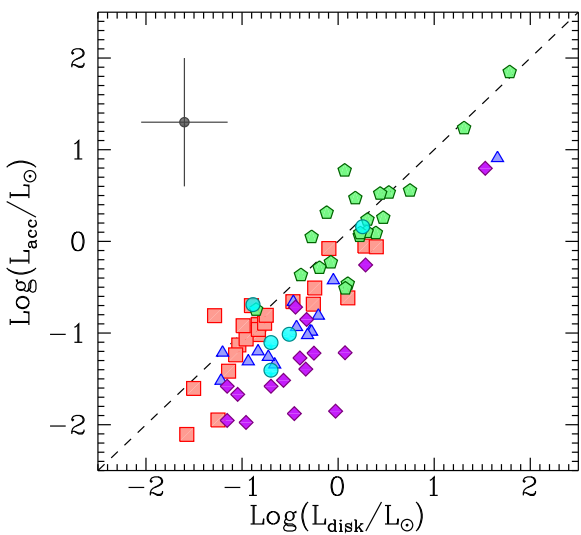}
\caption{\label{fig:acc_vs_ldisk} Accretion luminosity plotted as a function of the disk 
luminosity for the POISSON sources. Symbols are the same as in Fig.~\ref{fig:lacc_lstar_all}.
The dashed line shows the locus of equal \lacc\ and \ldisk.
}
\end{figure}

In Fig~\ref{fig:acc_vs_ldisk} we compare the accretion luminosities to the disk luminosity \ldisk\ of 
the objects, that is, \lbol\ -- \lstar, as defined in Sect.~\ref{sec:parameters} (details on the computation of \ldisk\ for 
L1641 and Cha objects are given in the appendix).
The plot shows that \ldisk\ $\sim$ \lacc\ for most of POISSON objects, which indicates that the luminosity in excess of the photosphere is basically provided 
by the accretion process, that is, from the shock caused by the disk material falling onto the star.
%
\nocite{nisini05a}  \nocite{antoniucci08}
However, the plot also shows that several sources, especially in the Lupus sample, 
display an accretion luminosity that is significantly lower (about one order of magnitude)
than the observed disk luminosity. 
Therefore in these stars the bulk of the excess luminosity does not come from the accretion process, which suggests intrinsic emission
from massive disks. 

Alternatively, a possible interpretation is that these might be edge-on objects in which 
the central star and innermost accretion region are heavily extincted 
(the so-called sub-luminous sources, see e.g. discussion in Alcal\`a et al. 2014), causing an 
underestimation of both \lacc\ and \lstar\ that is difficult to predict.

\subsection{Mass accretion rates}
\label{sec:macc}

\begin{figure*}[] 
\centering
\includegraphics[width=9.5cm]{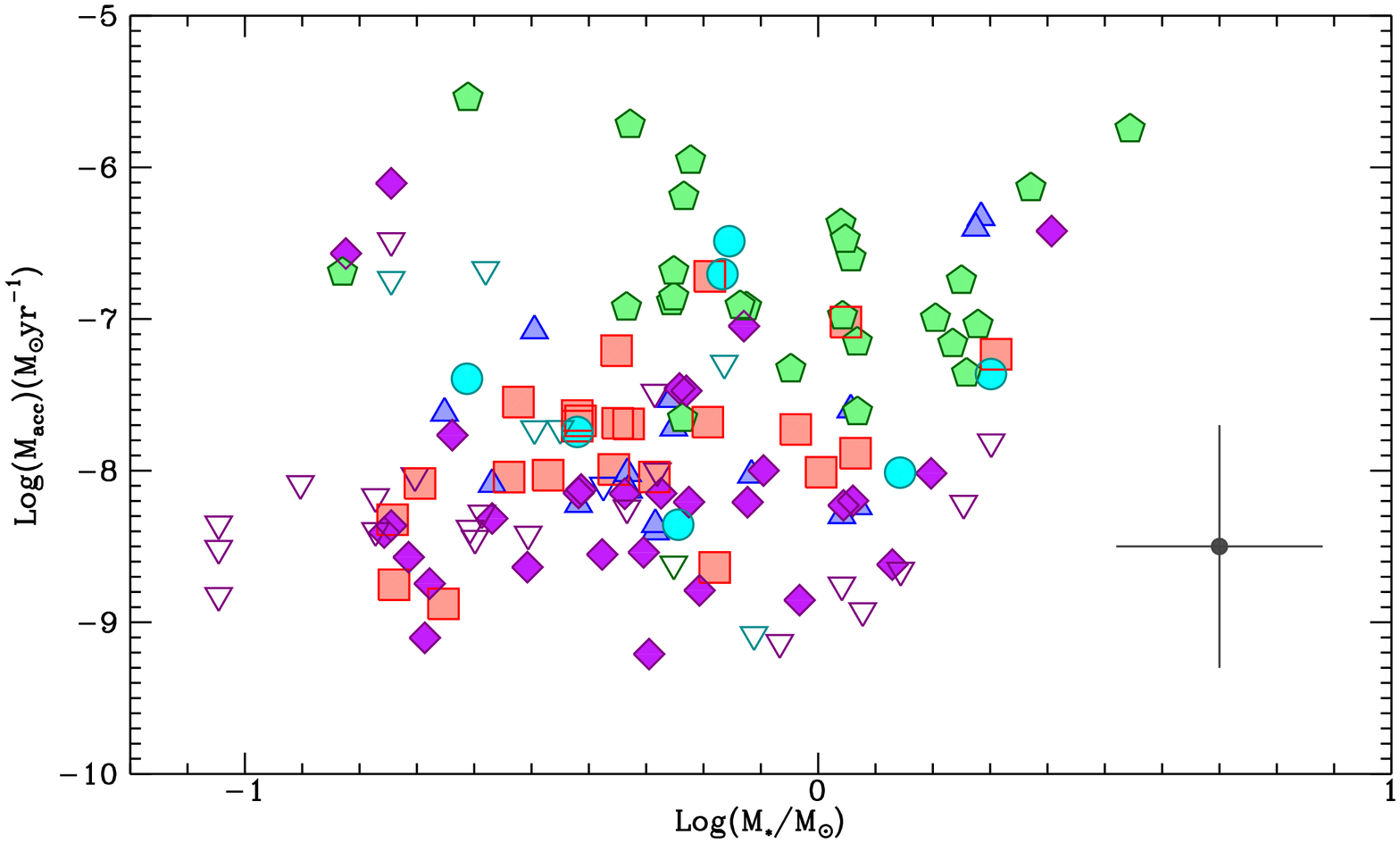}\hspace{-5ex}
\includegraphics[width=9.5cm]{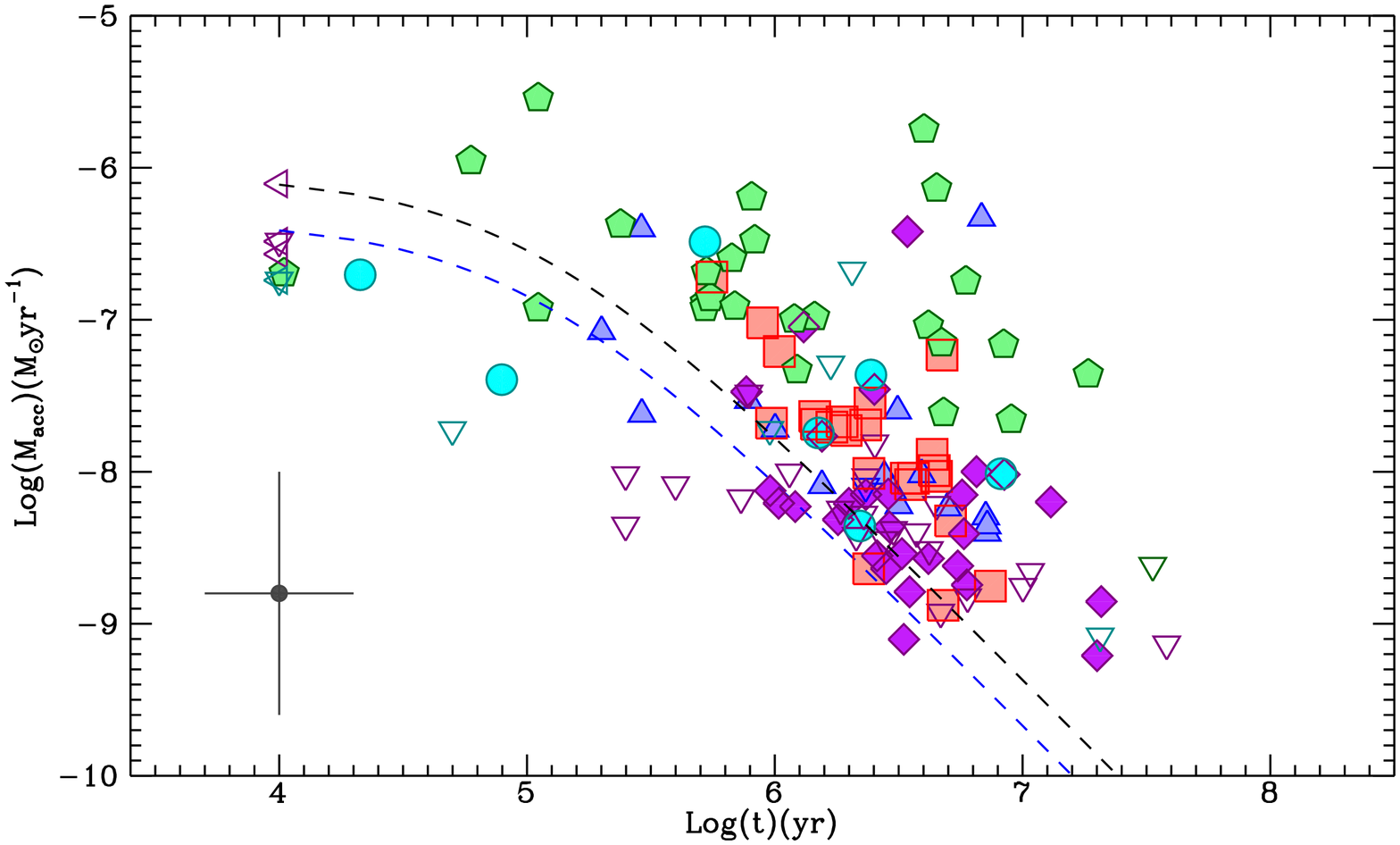}
\caption{\label{fig:macc_rels}  Mass accretion rates of the whole POISSON sample plotted 
as a function of the stellar mass (left) and age of the targets (right). Upper limits on \macc\ are indicated 
by downward triangles. The typical error bar of a single point is also shown.
The dashed lines in the right graph show the expected \macc\ evolution for a fiducial model of a viscous disk \citep{hartmann98} (see text for more details)
around a 0.5\msun\ star with initial disk masses equal to 0.1\msun\ (lower line) and 0.2\msun\ (upper line). 
Sample symbols are the same as in Fig.~\ref{fig:lacc_lstar_all}. 
}
\end{figure*}

The mass accretion rates of the whole sample range 
from $10^{-7}$ to $10^{-9}$~\msunyr, with a few objects in L1641 showing higher \macc\ values of 
about $10^{-6}$~\msunyr.
Based on the same considerations as in Sect.~\ref{sec:accretion_lup_ser}, we can assume 
a typical uncertainty on these \macc\ values of about 0.8 dex.
In Fig.~\ref{fig:macc_rels} we analyse the correlation of the mass accretion rate with both the mass and age 
of the POISSON sample (left and right panels).
Despite the huge scatter, a rough trend of \macc\ increasing with \mstar\ can be seen 
in the plot on the left. Indeed, from computing the Spearman rank correlation coefficient ($r$), we find $r=0.42$ with a probability of 
obtaining this from randomly distributed data $p \sim 10^{-5}$.

A power-law relationship \macc\ $\propto M_{*}^\beta$ with $\beta$ around 2 has been observed 
in several star-forming regions in recent years, for samples with masses and ages very similar 
to those explored by POISSON, see for instance $\beta$=2.1 \citep{muzerolle05},
$\beta$=1.8 \citep{natta06}, $\beta$=1.9 \citep{herczeg08},  
$\beta$=1.8 \citep{alcala14}. 
The function \macc\ $\sim M_{*}^{2}$ is marked with a dotted line in the left panel of Fig.~\ref{fig:macc_rels}
for reference.

In paper I, we found an analogous correlation ($\propto$\mstar$^{1.95}$) from limiting our analysis only 
to Cha~I objects. Here we considered all POISSON sources to improve the statistics, but found that 
the data points are so scattered (with a dispersion even greater than two orders of magnitude) that it is practically impossible
to achieve a reliable fit on the whole dataset.

A trend of accretion rate decreasing with time is visible for the \macc\ versus age plot.
The five sub-samples display different median values of \macc\ (disregarding upper limits), 
from $\sim 1 \times 10^{-7}$~\msunyr in L1641 down to $\sim 6 \times 10^{-9}$~\msunyr in Lupus. 
This decreasing median rate reflects an increasing median age of the 
five investigated samples, which spans from $\sim 8 \times 10^{5}$~yr 
for L1641 to $\sim$3~Myr for Lupus, in agreement with the expected decay of \macc\ as sources evolve.

The observed evolution trend can be compared with the \macc\ temporal variation 
predicted by \citet{hartmann98} for viscous disk models (see for example Muzerolle et al. 2000\nocite{muzerolle00}, 
Sicilia-Aguilar et al. 2010\nocite{sicilia-aguilar10}, and paper II).  
A typical fiducial model considered by \citet{hartmann98} assumes a 0.5 \msun\ star, constant 
viscosity $\alpha=10^{-2}$, and a viscosity exponent $\gamma=1$.
\macc\ evolution predictions based on this fiducial model are shown in Fig.~\ref{fig:macc_rels} 
using two dashed curves, which refer to two different initial disk masses of 
0.1 (lower) and 0.2 (upper) \msun.

\begin{figure*}[] 
\centering
\includegraphics[width=9.5cm]{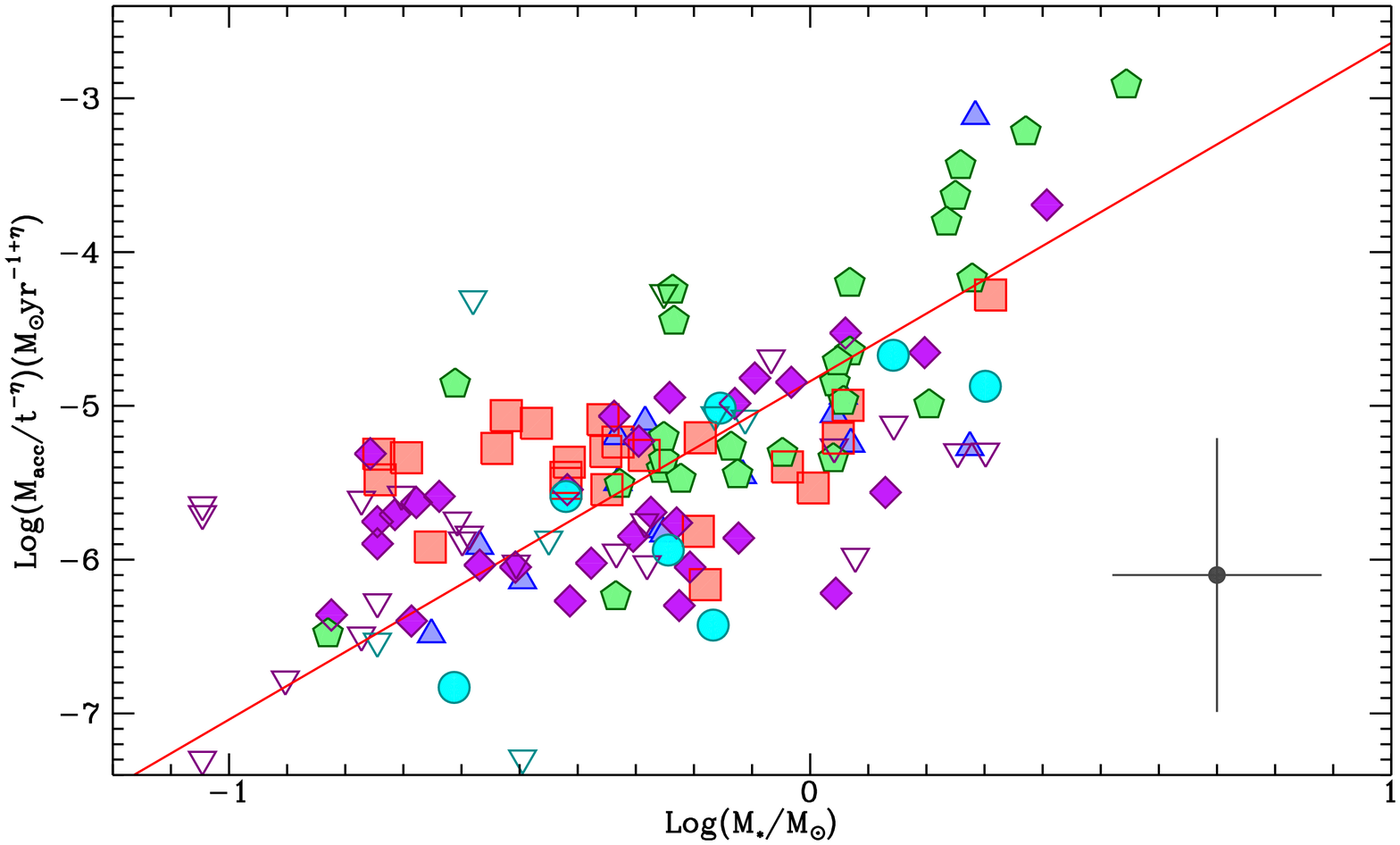}\hspace{-5ex}
\includegraphics[width=9.5cm]{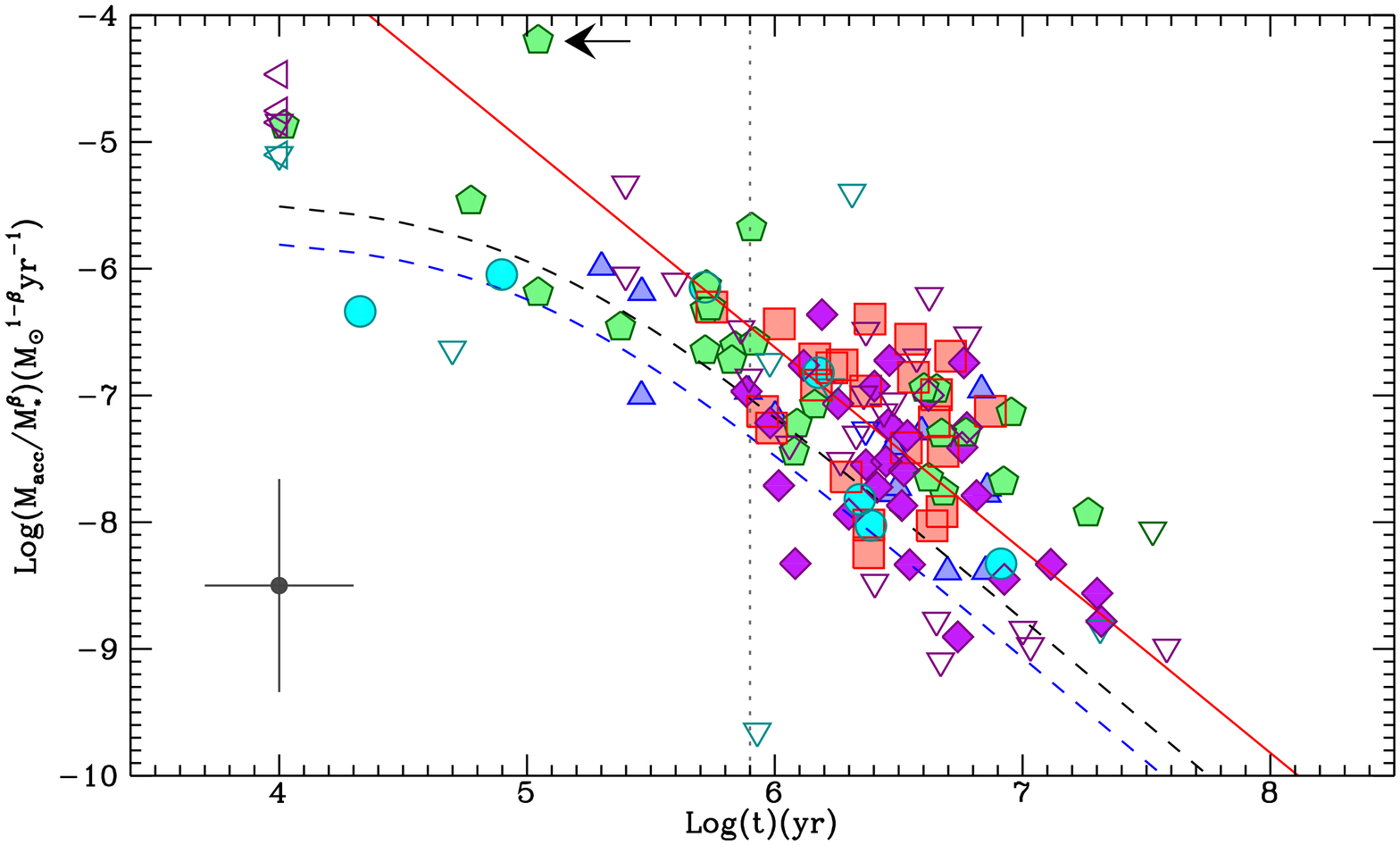}
\caption{\label{fig:norm_macc_rels} Same as in Fig.~\ref{fig:macc_rels}, but for normalised 
data: \macc\ normalised by the time factor provided by \citet{hartmann98} \macc\ evolution 
models ($\sim t^{-\eta}$, see text) vs \mstar\ (right) and 
\macc\ normalised by \mstar$^\beta$ vs age (right; \macc\ evolution curves by Hartmann et al. are also normalised). 
The best-fit relationships we obtain are indicated with a solid red line
and correspond to power-law indexes $\beta=2.2$ (left) and $\eta=1.6$ (right).
The vertical dotted line in the right graph shows the lower limit on age that we considered for the fit of the power -law decrement.
The position of the outbursting source V2775 Ori (paper II, Caratti o Garatti et al. 2011) is marked by an arrow in the
temporal evolution plot.
}
\end{figure*}

In these models the \macc\ evolution is typically slower at early time, then 
it asymptotically tends to a power-law decay of the type \macc\ $\sim t^{-\eta}$,
when the elapsed time is much longer than the initial characteristic viscous time of the disk.
For the fiducial models reported, the asymptotic relationship is \macc\ $\sim t^{-1.5}$, which 
is approximately reached when Log $t > 5.9$.
We also note that the initial disk mass ($M_D$) substantially sets the initial value of the accretion rate.
Accretion luminosities in the range 0.1~\lstar$<$\lacc$<$~\lstar, such as those observed by POISSON, are compatible 
with sources having a mass of 0.5-1.0~\msun\ and a disk-to-star mass ratio of 0.2-0.1 \citep{tilling08}, so that
the 0.1-0.2~\msun\ disk masses considered in the plot are in the range of the expected upper limit values of
$M_D$ for a solar-type star.
Even if the decay trend of accretion rates is very similar to the expected evolution for viscous 
disks, most of the measured \macc\ values appear to be higher than those predicted by the models.
Moreover, the data point spread at similar age is very large, so that we find several 
sources that show accretion rates up to two orders of magnitude greater than those of the fiducial viscous models
considered, in particular among L1641 objects. 

Before further investigating this discrepancy between models and observed points (Sect.~\ref{sec:discuss_macc}), 
we applied a procedure similar to the one we used in paper II for L1641 objects and normalised 
\macc\ by $M^\beta$ (when plotting \macc\ versus time) and $t^{-\eta}$ (when plotting \macc\ versus \mstar), 
which we can assume as the power-law index relationships that describe the dependence of 
\macc\ on \mstar\ and age\footnote{For the time normalisation we used the whole time term appearing in the 
model by \citet{hartmann98} (see their Equation 35); this temporal term becomes $\sim t^{-\eta}$ at long enough times.}.
In fact, since the accretion rate depends on both \mstar\ (explicitly, see Eq.~\ref{eq:macc}) and age 
(implicitly, because the accretion rate is expected to vary with time), this operation allows us to remove 
from the plotted accretion rate values the implicit dependence on one of the two parameters, so that
in the resulting plot we are able to better visualise the actual \macc\ dependence on the other 
parameter alone. In particular, normalising by the age and mass helps to reduce the impact on our analysis that arises from the
slightly different mass and age distributions of the samples. 

Because we do not know a priori the values of $\eta$ and $\beta$, we 
used a procedure in which we simultaneously fit the two normalised datasets 
to derive the best-fit values of both power-law indexes.
Using an ordinary least-square bisector linear regression\footnote{In our case this appears to be the best choice based on the fact that
we do not know the underlying functional relation between the plotted variables \citep[see][]{isobe90}} and excluding upper limits, 
we obtain the best-fit relations marked as solid red lines in Fig.~\ref{fig:norm_macc_rels}. 

For the dependence on the stellar mass alone we obtain \macc\ $\sim$ \mstar$^{2.2\pm0.3}$ with a significantly stronger higher coefficient ($r=0.62$, $p<10^{-10}$) than in the non-normalised plot. The power-law index is therefore consistent with the slope $\beta \sim 2$ observed in other star-forming regions.
\citet{ercolano14} have recently considered more than 3000 \macc\ determinations from the literature, obtained in various star-forming regions, to derive the power-law index of the \macc--\mstar\ relationship. The authors infer a value of $\beta$ in the range 1.6--1.9, but limited to stars with \mstar $<$ 1\msun.
Interestingly, if we restrict our fit to stars with masses below 1\msun\ in our sample, we obtain a power-law index of 1.7, fully consistent with the result found by Ercolano et al.

For the rate evolution at long times \macc\ decays as $t^{-1.6\pm0.2}$. 
We point out that in this case we limited the regression only to the region of the plot where 
the evolution is described by a power-law function (i.e. ages greater $t \gtrsim$ 1~Myr).
Also here we obtain a tighter (anti)correlation for normalised data, that is, $r=-0.42$ with $p\sim 0.0001$ against 
$r=-0.28$ with $p\sim 0.01$ for the previous plot.
We note that three sources in Lupus and one in L1641 show ages older than 10 Myr, which appears to be strangely high for these regions. As this might stem from 
an incorrect determination of the stellar luminosity (e.g. sources viewed edge-on), we decided to perform the fit also by excluding these objects. We still obtain an
accretion decay that behaves as $t^{-1.6\pm0.2}$, although with a weaker correlation ($r=-0.34$ with $p\sim 0.003$).
The value of the power-law index $\eta$ is very similar to that of the fiducial model considered 
($\eta$=1.5), and it falls within the interval 1.5-2.8 originally suggested by \citet{hartmann98}.
Thanks to the fairly large statistics available, we are confident that the observed general trend is real, even if uncertainties on age and accretion luminosities of the single points are relatively large. Nonetheless, most of the \macc\ values still appear to be higher
than those predicted for the viscous models: possible reasons for this finding are discussed in the following sections.

\begin{figure}[] 
\centering
~\\[-5ex]
\includegraphics[width=7.7cm]{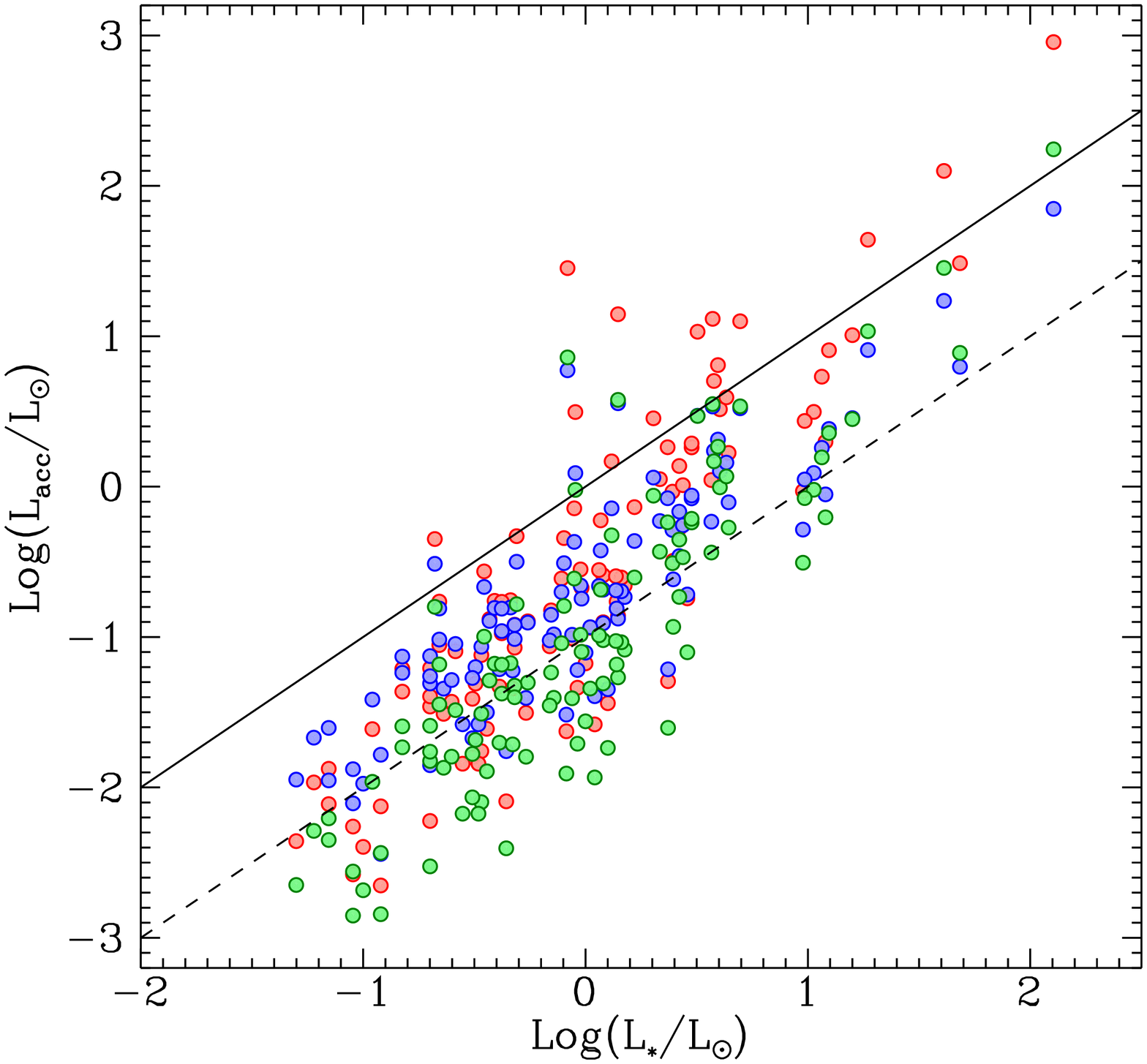}\\[-3ex]
\includegraphics[width=7.7cm]{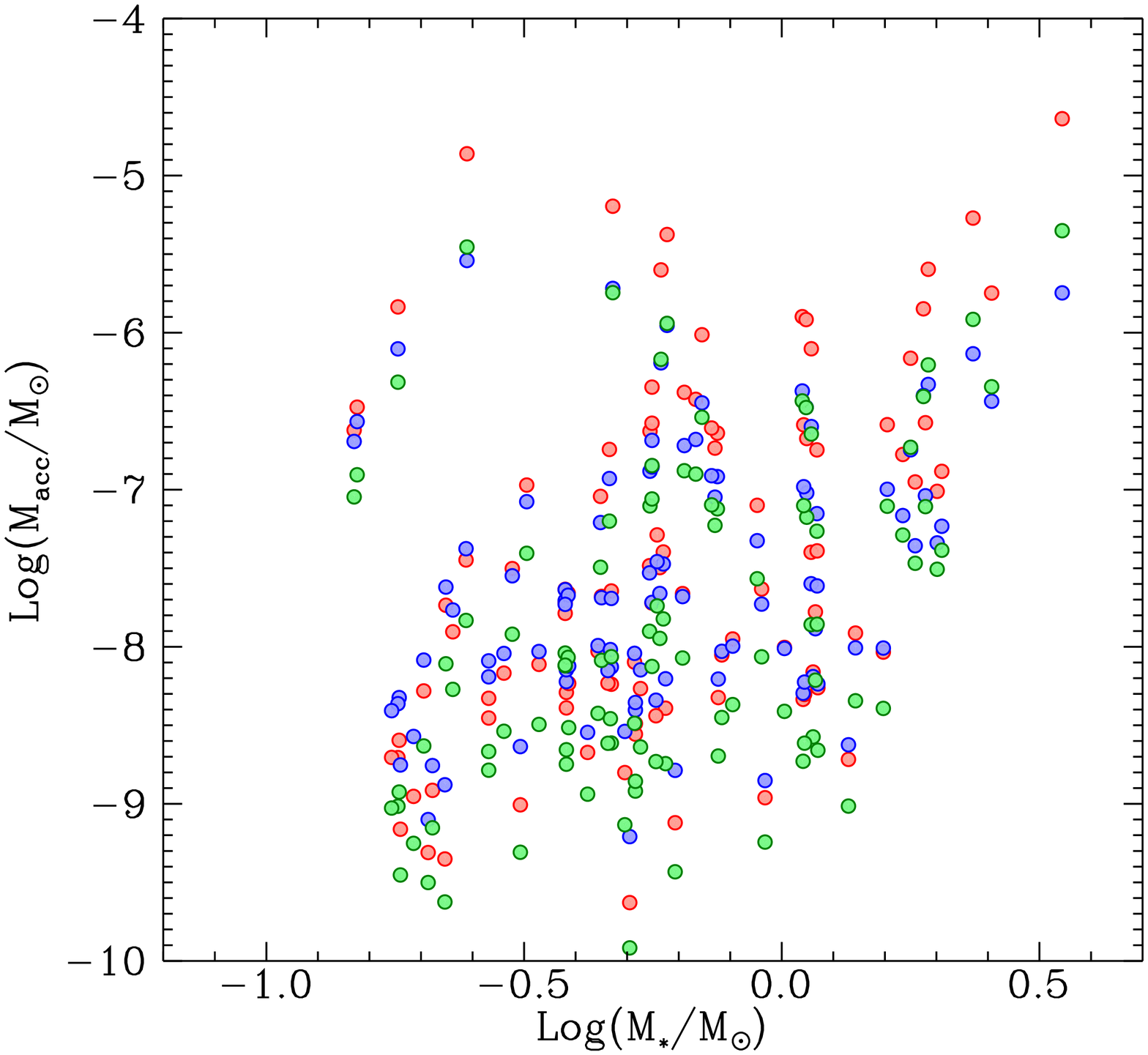}\\[-3ex]
\includegraphics[width=7.7cm]{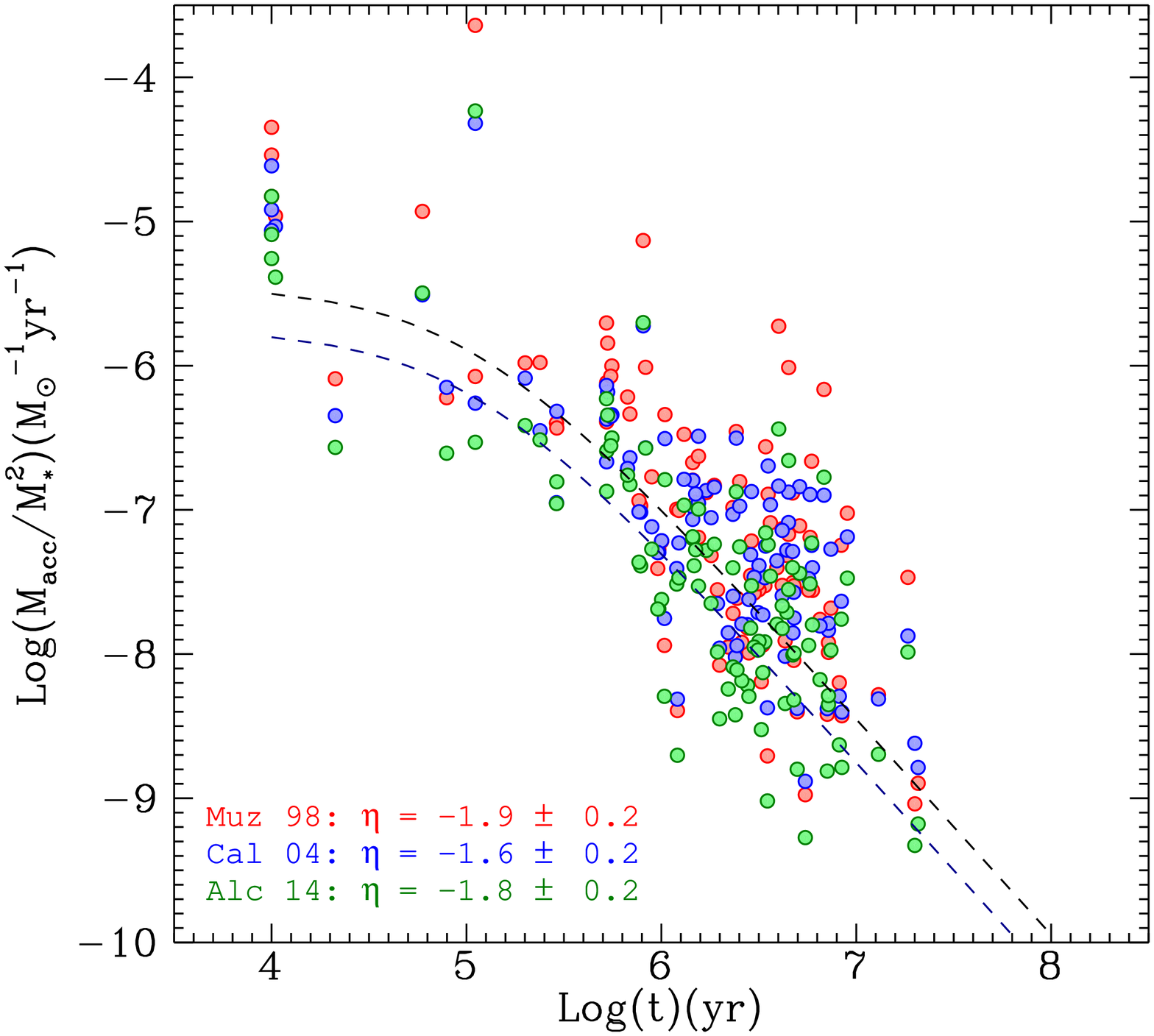}
\caption{\label{fig:relations} Results obtained by considering \lacc\ values from with three different 
empirical relationships for \brg: \citet{alcala14} (green), \citet{muzerolle98a} (red), and 
\citet{calvet04} (blue). Displayed plots are \lacc\ versus \lstar\ (top), \macc\ versus \mstar\ 
(centre), and \macc\ normalised to \mstar$^2$ versus age (bottom). In the last plot we report the curves of the viscous disk model of
\citet{hartmann98} shown in Figs.~\ref{fig:macc_rels} and \ref{fig:norm_macc_rels}; the best-fit
power-law indexes obtained from the different distributions for sources with Log $t > 5.9$ (see text for details) are also indicated.
}
\end{figure}

\subsection{Results from different line-\lacc\ relationships}
\label{sec:discuss_relations} 

In Sect.~\ref{sec:macc} we found that \macc\ values obtained from Eq.~\ref{eq:calv1} appear to be 
overestimated in general with respect 
to the predictions of viscous disk models. Therefore, we investigated 
the possible systematic effects caused by the empirical relationships adopted to
convert \hi\ line luminosities into \lacc.
We considered two additional relationships, the classical formula provided by \citet{muzerolle98a} 
and the recent calibration by \citet{alcala14}, based on VLT/X-Shooter observations of a sample of 
Lupus objects:
\begin{equation}
\label{eq:muzerolle}
\mathrm{Log}\;L_{acc}/L_\odot = 1.26 \cdot \mathrm{Log}\;L_{\mathrm{Br\gamma}}/L_\odot + 4.43 ~,
\end{equation}
\begin{equation}
\label{eq:alcala}
\mathrm{Log}\;L_{acc}/L_\odot = 1.16 \cdot \mathrm{Log}\;L_{\mathrm{Br\gamma}}/L_\odot + 3.60 ~.
\end{equation}
The three relationships yield accretion luminosities that may significantly differ from each 
other depending on the line luminosity range involved. This is evident in the upper panel of 
Fig.~\ref{fig:relations} where we show the \lacc\ vs \lstar\ plot we obtain with the three relationships. 
The different \lacc\ vs \lstar\ trends we observe 
obviously reflect the different parameters of the laws, and the relationships of Muzerolle and Alcal\'a provide 
lower accretion luminosities for lower luminosity stars. In particular, the differences with 
values reported by Alcal\'a are about 0.7 dex for the less luminous stars of our sample ($\sim$0.4 dex for a solar-type star).

Of course, the same differences are found for the mass accretion rates, which are directly 
proportional to \lacc.
The \macc\ vs \mstar\ plot is shown in the central panel of Fig.~\ref{fig:relations}. 
In this case, we note that although the single \macc\ values may significantly differ in the three cases,  
the general distribution of the points in the plot does not change substantially.

The \macc\ evolution plots displayed in the lower panels of Fig.~\ref{fig:relations} show that 
the generally lower \macc\ values obtained from the Alcal\`a 
(and Muzerolle) relationship agree better with the rates expected for a viscous 
disk evolution, with more points falling along or below the viscous disk curves of the 
reference fiducial models we considered in Sect.~\ref{sec:macc}.
However, even though the general agreement of the measurements with the models is improved, 
we still observe a large number of source points located above the curves. This means that 
the systematic effect caused by the adopted relationship alone cannot explain these higher-than-expected \macc\ values, 
so we need to invoke other causes, which we discuss in Sect.~\ref{sec:discuss_macc}.

Additionally, although the use of a different line-\lacc\ relation leads to fairly 
large variations of the single accretion rate estimates, this does not seem to heavily affect the 
general trend of the \macc\ evolution. Indeed, we obtain that the power-law index describing 
the \macc\ time evolution is very similar in the three cases, with best-fit $\eta$ values 
of 1.6$\pm$0.2 (Calvet), 1.9$\pm$0.2 (Muzerolle), and 1.8$\pm$0.2 (Alcal\'a), which are 
substantially consistent with each other considering uncertainties (see Fig~\ref{fig:relations}).

\subsection{Accretion rate evolution}
\label{sec:discuss_macc} 

In the previous sections we found that the variation of \macc\ with time roughly follows the  
trend expected from viscous evolution, although two open problems needed to be further investigated: 
\textit{i)} the evidence of a large spread of \macc\ values for objects with similar age, even after considering
"normalised" accretion rates, and \textit{ii)} the presence of many measured accretion rates significantly higher than those expected 
for a viscous disk evolution \citep{hartmann98}, which we obtain also allowing for systematic effects due to the
employed empirical \hi-\lacc\ relation (Sect.~\ref{sec:discuss_relations}).

The time evolution of the accretion rate has recently been studied in several papers
\citep[e.g.][]{muzerolle00,sicilia-aguilar06,sicilia-aguilar10}. In particular, \citet{sicilia-aguilar10} analysed
the relationship between accretion rates and ages collected for sources from various star-forming 
regions (see their Fig.~2).
These authors found a good general agreement between \macc\ measurements and the evolution trend predicted by viscous
disk models, although in their analysis the measurements are also characterised by a large spread
around the model curves and by numerous data points that are higher than those predicted for fairly massive 
accretion disks.

Despite the uncertainties, the large \macc\ spread we observe is most likely a real effect. 
Indeed, within the framework of viscous accretion, a high dispersion of \macc\ is automatically
obtained by considering a range of initial disk masses 
because the predicted \macc\ is directly proportional to $M_D$ \citep{hartmann98}. 
This scenario can partly explain the dispersion observed, but it can be reasonably 
invoked only for points that lie around or below the curves relative to $M_D$
of 0.1 and 0.2 \msun (see Fig.~\ref{fig:norm_macc_rels}), which substantially represent 
the upper limit for disk masses around solar-type stars.

Another possible way to explain the scatter of \macc\ points is to assume that the initial conditions and/or the physical processes are not the same in the various disks. 
Strong observational evidence supporting the presence of very different viscosity laws in the disks was recently 
provided by \citet{isella09}, who revealed a wide variety of viscosity parameters
for several solar-type objects.
For instance, by considering a different viscosity parameter $\alpha$ in the disk models, we obtain
\macc\ evolution curves that may significantly differ from the fiducial model curves considered in 
Sect.~\ref{sec:macc} \citep[see examples in][]{hartmann98}.

Varying accretion through recurrent bursts may also be at the origin of the strong discrepancies of \macc\ 
with respect to the expected average level of accretion at a certain age.
Powerful accretion bursts are predicted by recent accretion theoretical models \citep[e.g.][]{vorobyov06}.
Although we have evidence that most sources of the sample are not in an outbursting stage, variable accretion may 
effectively contribute to the observed scatter of \macc\ data points.
In our sample we find clear evidence for this effect in [MJS2008] 146 and especially 
in the extreme case of V2775 Ori, which shows an \macc\ about two orders of magnitude greater than predicted 
in the normalised plot of Fig.~\ref{fig:norm_macc_rels}.

While for the accretion rate decay at long times we obtain \macc\ $\sim t^{-1.6}$, which is very similar 
to the $\sim t^{-1.5}$ trend of the fiducial viscous disk model, \citet{sicilia-aguilar10} found a relationship \macc\ $\sim t^{-1.2}$,
suggesting a somewhat slower decrease of the mass accretion rate.
Noticeably, this is the same value of the power-law index as we inferred for L1641 objects alone in paper II, although
we point out that in our previous work we considered also younger sources in the fit, with ages between 
10$^5$ and 10$^6$ yr, in which the asymptotic \macc\ evolution trend might not be reached yet.
Our analysis is limited because the bulk of the POISSON data points is in the (restricted) age range of 1-10 Myr and 
typically presents an \macc\ dispersion of about $\sim$2 dex.
The sample studied by \citet{sicilia-aguilar10} has the advantage of presenting a mean age older than the regions studied here, 
which allows on the one hand to better constrain the \macc\ decay for t $\gtrsim$ 10 Myr, 
and on the other hand to somewhat minimise the age spread uncertainties for individual objects.

On the other side of the plot, the very low number of very young sources allows us to derive only a marginal evidence 
of the expected slower evolution of the mass accretion rate at early times ($t \lesssim$ 1Myr) in 
Fig.~\ref{fig:norm_macc_rels}.
The moment of the onset of the asymptotic decline of \macc\ depends on the parameters of the disk
(such as viscosity and dimensions), therefore a better sampling of this region would possibly
provide information on these aspects.

Another effect that may alter the evolution of the mass accretion rate is the presence
of additional processes that might be adding to disk dissipation, such as 
photo-evaporation \citep[e.g.][]{gorti09, owen12}.
Indeed, a very efficient photo-evaporation would dissipate the disk in a short time because 
a rapid cut-off of \macc\ (with a time-scale of about 10\% of the disk lifetime) is expected when the accretion rate 
drops below the photo-evaporation rate \citep[e.g.][]{ercolano14}. 
\citet{sicilia-aguilar10} invoked this mechanism to explain the presence of many objects without 
significant accretion (\macc$\lesssim 10^{-11}$~\msunyr) in a wide range of ages ($t \gtrsim 1$Myr).
In our sample we observe only a few of these upper limits in Fig.~\ref{fig:norm_macc_rels}, 
mostly among evolved Lupus sources with ages $>$5Myr.
This finding may suggest that the photo-evaporation is not the dominant process in many of the disks observed by POISSON,
although its presence can accelerate the decay of \macc\ in some objects and thus produce the large spread of the accretion rates that we observe.
Indeed, the models \citep[e.g.][]{owen11, owen12} show that the quick drop of \macc\ at times $t \gtrsim 1$Myr induced by the X-ray photo-evaporation 
may easily produce a large spread of the accretion rates starting from the expected viscous decay,
depending on the X-ray flux of the sources (see for instance Fig.~7 of Owen et al., 2011) and on the initial disk properties.
Moreover, this would provide a somewhat steeper general decline of \macc\ with respect to viscous evolution when fitting the whole sample, which we marginally detect.
In a scenario where photo-evaporation regulates disk dissipation for most objects, the lack of detection of non-accreting sources might be related to the 
the selected sample being only composed of relatively bright objects, 
which might partly explain why our results appear generally biased towards high 
average values of the mass accretion rate.

\section{Conclusions}
\label{sec:conclusions}

We have presented the results of POISSON near-infrared spectroscopic observations of two samples of low-mass (0.1-2.0\msun) young stellar 
objects in Lupus (52 sources) and Serpens (17 sources). 
We have derived the accretion luminosities \lacc\ and mass accretion rates \macc\ of the targets using the 
\hi\ \brg\ line (\pab\ in a few cases) and the relevant empirical relationship connecting the flux of the line 
to \lacc.

The new results from the Lupus and Serpens samples were then added to previous results from 
our POISSON papers I and II (on Cha~I,Cha~II, and L1641 clouds) to build a large catalogue of the
whole POISSON sample, composed of 143 objects with masses in the range 0.1-3.5 \msun\ and ages between $10^4$ and few $10^7$ years.
For all these sources we obtained \macc\ estimates computed in a homogeneous and consistent fashion and analysed
in particular the \macc\ correlation with stellar mass (\mstar) and its evolution through time.

We observe a correlation \macc$\sim$\mstar$^{2.2}$ between mass accretion rate and stellar mass, similar 
to the correlation observed in several star-forming regions. 
We also find that the temporal variation of \macc\ is roughly consistent with the expected evolution of the accretion rate in 
viscous disks, with an asymptotic decay proportional to $t^{-1.6}$. However, sources with similar age are characterised by 
\macc\ values that display a large scatter and are generally higher than the predictions of viscous models. 
Although part of these aspects may be related to systematics due to employed empirical relationship and to the relatively big 
uncertainties on the single measurements, the large statistics available make us confident that the general distribution and decay 
trend of the \macc\ points are real.
In particular, the observed \macc\ scatter might be indicative of a large variation in the initial mass of the disks, 
of fairly different viscous laws among disks, of varying accretion 
regimes, and of other mechanisms that add to the dissipation of the disks, 
such as photo-evaporation.

\section*{Acknowledgements}
\begin{footnotesize}
We are very grateful to the anonymous referee for all suggestions and comments, which helped us to improve the quality of the paper.
The authors wish to thank J. M. Alcal\'a for fruitful discussions and L. Siess for providing help for 
the interpolation on the grid of pre-main sequence evolutionary models.\\
The authors acknowledge the funding support from the PRIN INAF 2012 ``Disks, jets, and the dawn of planets".
\end{footnotesize}

\bibliographystyle{aa} 
\bibliography{refs} 


\section*{Appendix A: near-IR photometry.}
\label{apx:photometry}

Measured $JHK$ magnitudes of the sources are reported in Tables~\ref{tab:phot1} and \ref{tab:phot2} together with 2MASS photometry, which is given for comparison.
Small photometric variations (over time-scales of months/years) of about a few tenths of magnitude in the near-IR are typical of young objects 
\citep[e.g.][]{alves_de_oliveira08} and are generally attributed to varying accretion activity or extinction variations (see e.g. Carpenter et al. 2001). 

In our samples most of the sources display this type of fluctuation, with a mean magnitude 
variation (in absolute value) of about 0.2, 0.1, and 0.1 mag ($JHK$-bands, respectively) in Lupus and 
0.2, 0.2, and 0.4 mag in Serpens.

However, we were also able to identify a few objects that underwent significant brightness 
variations (of about 1 mag or greater in at least one of the bands): Sz98, [MJS2008]~133, 
and [MJS2008]~146 in Lupus and [WMW2007]70 and [WMW2007]4 in Serpens.
In particular, [MJS2008]~133 appears much dimmer in POISSON data than in 2MASS observations and displays 
photometric variations of $\Delta J=-1.5$, $\Delta H=-1.9$ $\Delta K=-2.0$ mag \citep{antoniucci13a}. 
These pronounced photometric fluctuations might indicate an EXor- (e.g. Lorenzetti et al. 2011) or 
UXor-type \citep[e.g.][]{shenavrin12} variability for these objects. 

\begin{table}[]
\caption{Near-infrared photometry of the Lupus targets derived from POISSON data, compared with 2MASS magnitudes. 
Typical uncertainty on the provided values is 0.05 mag for the $H$ band and 0.1 mag for $J$ and $K$ bands (see text for details).}
\label{tab:phot1}
\begin{tiny}
\begin{tabular}{l|ccc|ccc}
\hline
\hline
ID  &  $J$  &  $H$  &  $K$ &  $J$  &  $H$  &  $K$  \\
  &  \multicolumn{3}{c|}{2MASS}  &  \multicolumn{3}{c}{POISSON}  \\
\hline
 01     &  9.19  &  8.41  &  7.98  &  8.9  &  8.32  &  8.1  \\ 
 02     &  10.89  &  9.88  &  9.29  &  10.2  &  9.54  &  9.2  \\ 
 03     &  12.2  &  10.64  &  9.71  &  11.9  &  10.53  &  9.8  \\ 
 04     &  7.57  &  6.86  &  6.48  &  7.2  &  6.81  &  6.5  \\ 
 05     &  11.18  &  10.16  &  9.41  &  10.8  &  10.02  &  9.42  \\ 
 06     &  11.64  &  10.95  &  10.48  &  11.8  &  11.15  &  10.4  \\ 
 07     &  10.07  &  9.18  &  8.63  &  10.2  &  9.4  &  9.0  \\ 
 08     &  10.57  &  9.77  &  9.33  &  10.7  &  9.93  &  9.5  5\\ 
 09     &  10.74  &  9.53  &  8.83  &  10.8  &  9.59  &  8.7  6\\ 
 10     &  8.69  &  7.7  &  7.1  &  8.3  &  7.41  &  6.8  \\  
 11     &  8.78  &  8.09  &  7.74  &  8.6  &  7.87  &  7.5  \\ 
 12     &  8.73  &  7.82  &  7.14  &  8.6  &  7.74  &  7.0  \\ 
 13     &  10.93  &  10.2  &  9.85  &  11.0  &  10.27  &  10.0  \\ 
 14     &  8.55  &  7.69  &  6.98  &  8.2  &  7.59  &  7.0  \\ 
 15     &  8.16  &  7.06  &  6.12  &  7.3  &  6.33  &  5.8  \\ 
 16     &  8.71  &  7.53  &  6.9  &  8.8  &  7.92  &  7.3  \\ 
 17     &  9.46  &  8.68  &  8.35  &  8.9  &  8.38  &  8.2  \\ 
 18     &  9.73  &  8.96  &  8.5  &  9.6  &  8.98  &  8.8  \\ 
 19     &  12.11  &  10.55  &  9.53  &  11.8  &  10.43  &  9.5  \\ 
 20     &  9.92  &  9.12  &  8.56  &  9.8  &  9.1  &  8.6  \\ 
 21$^a$ & ...  &  ...  &  ...  &  11.3  &  10.7  &  10.4  \\ 
 22     &  11.51  &  10.65  &  10.11  &  11.4  &  10.68  &  10.2  \\ 
 23     &  10.35  &  9.32  &  8.72  &  10.5  &  9.48  &  8.9  \\ 
 24     &  11.01  &  10.28  &  10.01  &  11.0  &  10.27  &  10.1  \\ 
 25     &  10.03  &  8.53  &  7.67  &  9.7  &  8.33  &  7.6  \\ 
 26     &  10.13  &  9.35  &  8.96  &  10.3  &  9.46  &  9.0  \\ 
 27     &  11.24  &  10.55  &  10.22  &  11.2  &  10.55  &  10.3  \\ 
 28     &  9.53  &  8.65  &  8.01  &  10.7  &  9.47  &  8.6  \\ 
 29     &  11.93  &  11.21  &  10.75  &  11.8  &  11.22  &  10.9 \\ 
 30     &  10.98  &  10.35  &  9.91  &  10.8  &  10.29  &  9.9 \\ 
 31     &  11.38  &  10.62  &  10.23  &  11.1  &  10.59  &  10.4  \\ 
 32     &  8.97  &  8.39  &  8.22  &  9.0  &  8.38  &  8.2  \\ 
 33     &  11.67  &  11  &  10.65  &  11.5  &  10.98  &  10.7  \\ 
 34$^b$ &  5.91  &  5.22  &  4.39  &  6.0  &  5.22$^b$  &  4.3  \\ 
 35     &  11.65  &  10.66  &  10.15  &  11.5  &  10.53  &  10.1  \\ 
 36     &  11.25  &  10.62  &  10.31  &  11.1  &  10.58  &  10.4  \\ 
 37     &  11.4  &  10.82  &  10.5  &  11.3 &  10.82  &  10.7  \\ 
 38     &  11.45  &  10.17  &  9.55  &  11.5  &  10.27  &  9.8  \\ 
 39     &  10.97  &  10.22  &  9.75  &  10.8  &  10.17  &  9.9  \\ 
 40     &  11.33  &  10.28  &  9.8  &  11.3  &  10.28  &  9.9  \\ 
 41     &  10.62  &  9.8  &  9.54  &  10.5  &  9.82  &  9.6  \\ 
 42     &  11  &  10.29  &  9.96  &  11.0  &  10.32  &  9.9  \\ 
 43     &  12.47  &  11.72  &  11.26  &  12.5  &  11.64  &  11.1 \\ 
 44     &  10.41  &  9.7  &  9.32  &  10.3  &  9.64  &  9.3  \\ 
 45     &  10.68  &  9.85  &  9.43  &  10.5  &  9.84  &  9.7  \\ 
 46     &  10.45  &  9.35  &  8.68  &  10.4  &  9.38  &  8.8  \\ 
 47     &  12.66  &  12.09  &  11.76  &  12.8  &  12.11  &  11.8  \\ 
 48     &  13.28  &  12.65  &  12.32  &  13.4  &  12.72  &  12.3  \\ 
 49     &  11.09  &  10.21  &  9.78  &  11.0  &  10.18  &  9.8  \\ 
 50     &  11.84  &  10.28  &  9.41  &  13.3  &  12.19  &  11.4  \\ 
 51     &  10.63  &  9.6  &  8.96  &  10.5  &  9.5  &  8.9  \\ 
 52     &  10.54  &  9.77  &  9.54  &  10.5  &  9.82  &  9.5  \\ 
\hline
\end{tabular}
\end{tiny}
~\\
Notes.\\
$^a$ this secondary component was not spatially resolved in 2MASS. \\
$^b$ 2MASS $H$ magnitude was used to calibrate the RG spectrum of HR5999.
\end{table}

\begin{table}[]
\caption{Near-infrared photometry of the Serpens targets derived from POISSON data, compared with 2MASS magnitudes. 
Typical uncertainty on the provided values is 0.05 mag for the $H$ band and 0.1 mag for $J$ and $K$ bands (see text for details).}
\label{tab:phot2}
\begin{tiny}
\begin{tabular}{l|ccc|ccc}
\hline
\hline
ID &  $J$  &  $H$  &  $K$ &  $J$  &  $H$  &  $K$  \\
  &  \multicolumn{3}{c|}{2MASS}  &  \multicolumn{3}{c}{POISSON}  \\
\hline
01  &   11.6   &  10.39  &  9.64   &  ...      &  10.44  &   9.5  \\ 
02  &   13.9   &  12.45  &  11.6   &  ...      &  13.12  &  12.2  \\ 
03  &   16.89  &  15.66  &  12.67  &  ...      &  15.66  &  13.4  \\ 
04  &   12.61  &  11.24  &  10.41  &  12.7  &  11.38  &  10.6  \\ 
05  &   15.34  &  12.47  &  10.65  &  ...      &  12.86  &  11.3   \\ 
06  &   15.01  &  12.81  &  11.52  &  ...      &  12.65  &  11.6   \\ 
07  &   15.17  &  12.47  &  11.09  &  ...      &  12.46  &  11.2  \\ 
08  &   12.22  &  9.24   &  7.05   &  12.7   &  9.57   &   7.5 \\ 
09  &   16.21  &  12.79  &  10.81  &  ...      &  12.79  &  11.1  \\ 
10  &   18.17  &  15.82  &  12.97  &  ...      &  15.63  &  14.5  \\ 
11  &   17.2   &  14.41  &  11.66  &  ...      &  14.36  &  11.5  \\ 
12  &   15.75  &  12.46  &  10.52  &  ...      &  12.32  &  10.1  \\ 
13  &   13.07  &  11.19  &  9.86   &  ...      &  10.91  &   9.8  \\ 
14  &   15.19  &  14.03  &  11.84  &  ...      &  14.56  &  12.8  \\ 
15  &   15.2   &  11.92  &  9.92   &  ...      &  12.29  &  10.3  \\ 
16  &   12.28  &  11.05  &  10.38  &  12.3  &  10.96  &  10.4  \\ 
17  &   12.25  &  10.84  &  10.07  &  12.7   &  11.19  &  10.5  \\
\hline
\end{tabular}
\end{tiny}
\end{table}

\begin{figure}[!b] 
\centering
\includegraphics[width=9cm]{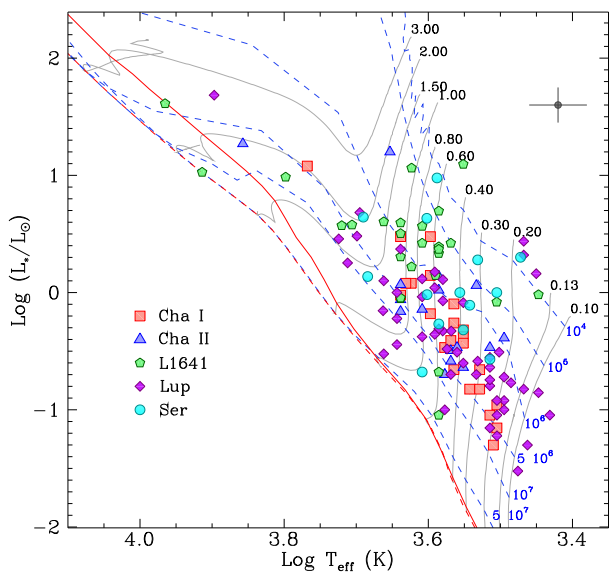}
\caption{\label{fig:diag} HR diagram for the sources of the five star-forming clouds analysed in POISSON (see legend). 
Superposed evolutionary tracks for labelled stellar masses in \msun\ (grey lines), isochrones for labelled time in years (blue dashed lines), ZAMS (solid red line), and early main 
sequence (dashed red lines) are taken from the models of \citet{siess00}. The point with error bars shows the mean uncertainty on the position of the single sources.
}
\end{figure}

\section*{Appendix B: determination of \ldisk, \mstar, and age.}
\label{sec:appendix}

Source disk luminosities employed in Sect.~\ref{sec:lacc} were not available in the literature for 
Cha (paper I) and L1641 objects (paper II), therefore these were obtained 
using the estimates of the bolometric and stellar luminosities: \ldisk=\lbol--\lstar.
However, since the \lstar\ values given in paper II for L1641 objects had been obtained as \lbol--\lacc,
to determine \ldisk\ entirely independently of \lacc\ 
(like for other sub-samples), we have re-derived \lstar\ from the spectral type and $I$ magnitude 
of the objects ($J$ magnitude when $I$ was not available) by
integrating a NEXTGEN stellar spectrum \citep{nextgen} of the same spectral type as the object normalised to the
observed $I$ ($J$) band flux. 
The new \lstar\ values thus obtained for L1641 are listed in Table~\ref{tab:hr1}.

As mentioned in Sect.\ref{sec:global_sample}, to minimise systematic biases 
we decided to (re-)determine \mstar\ and age of all POISSON sources on the basis
of the same pre-main sequence evolutionary models.
For this task we considered the models by \citet{siess00,siess99}, assuming a metallicity Z=0.2 
and no overshooting. The relative evolutionary tracks and isochrones are displayed in the 
HR diagram in Fig.~\ref{fig:diag}.
Mass and age of the single objects were computed by interpolating (where possible) over 
the model grid. For sources located outside the area covered by the model grid only 
upper limits to the parameters are provided. 

The main stellar parameters (\lstar, \teff, Spectral Type, \mstar, age) of all POISSON objects 
are listed (per region) in Tables~\ref{tab:hr1} and ~\ref{tab:hr2}, together with the 
determined mass accretion rate.

\begin{table*}[!t]
\caption{\label{tab:hr1} Main parameters of POISSON sources in Cha I, Cha II, and L1641, investigated in papers I and II. 
Masses and ages (re)determined from HR diagram position using the models of \citet{siess00}, in conjunction with \macc values derived from \brg.}
\begin{tiny}
\vspace{-.3cm}
\begin{tabular}{clcccccc}
\hline
\hline
ID  &  Name  &  \lstar(\lsun)  &  \teff(K)  & ST &  \mstar(\msun) &  age(yr) & \macc(\msunyr) \\ 
\hline
       \multicolumn{8}{c}{\bf{Chamaeleon I}} \\ 
\hline
01 & T11      &  1.20  & 4205  & K6 & 0.91 & 1.9E+06  & 1.9E-08    \\       
02 & CHSM1715 &0.05  & 3234  & M5 & 0.18 & 7.4E+06  & 1.8E-09    \\       
03 & T14      & 0.95  & 4350  & K5 & 1.16 & 4.3E+06  & 1.3E-08    \\       
04 & ISO52    &0.09  & 3270  & M4 & 0.22 & 4.8E+06  & 1.3E-09    \\       
05 & Hn5      & 0.11  & 3198  & M5 & 0.20 & 3.5E+06  & 8.2E-09    \\       
06 & ISO92    & ...    & ...   & ...& ...   & ...        & ...            \\       
07 & ISO97    & ...    & ...   & ...& ...   & ...        & ...            \\       
08 & T26      &  12.0  & 5860  & G2 & 2.04 & 4.7E+06  & 5.8E-08    \\       
09 & B35      & ...    & ...   & ...&  ...  & ...        & ...            \\       
10 & CHXR30B  & 0.22  & 3669  & M1 & 0.44 & 4.4E+06  & 1.0E-08    \\       
11 & T30      & 0.15  & 3488  & M2 & 0.34 & 4.5E+06  & 9.3E-09    \\       
12 & CHXR30A  &  1.40  & 3955  & K7 & 0.64 & 9.7E+05  & 2.9E-08    \\       
13 & T31      &  3.00  & 3955  & K7 & 0.65 & 5.6E+05  & 1.9E-07    \\       
14 & Cha IRN  &  ...   & ...   & ...& ...   & ...        & ...            \\       
15 & C9-3     &  ...   & ...   & ...& ...   & ...        & ...            \\       
16 & T38      & 0.34  & 3778  & M0 & 0.52 & 3.4E+06  & 9.1E-09    \\       
17 & CHXR79   & 0.55  & 3669  & M1 & 0.45 & 1.5E+06  & 2.0E-08    \\       
18 & C1-6     & 0.80  & 3669  & M1 & 0.45 & 1.0E+06  & 6.2E-08    \\       
19 & C1-25    &  ...   &...    & ...& ...   & ...        & ...            \\       
20 & Hn10-e   & 0.15  & 3379  & M3 & 0.29 & 3.6E+06  & 9.1E-09    \\       
21 & B43      & 0.22  & 3379  & M3 & 0.30 & 2.4E+06  & 2.8E-08    \\       
22 & T42      &  3.00  & 4350  & K5 & 1.12 & 8.9E+05  & 9.5E-08    \\       
23 & T43      & 0.48  & 3560  & M2 & 0.38  & 1.4e+06   & 2.3E-08    \\       
24 & C1-2     &  ...   &...    & ...& ...   & ...        & ...            \\       
25 & Hn11     & 0.66  & 3955  & K7 & 0.66 & 2.4E+06  & 2.3E-09$^a$   \\       
26 & ISO237   &  1.20  & 4278  & K5 & 1.01 & 2.4E+06  & 9.7E-09    \\       
27 & T47      & 0.42  & 3560  & M2 & 0.38  & 1.7E+06    & 2.0E-08    \\       
28 & ISO256   &0.07  & 3198  & M5 & 0.18 & 5.1E+06  & 4.7E-09    \\       
29 & T49      & 0.37  & 3560  & M2 & 0.38 & 1.9E+06  & 2.1E-08    \\       
30 & T53      & 0.39  & 3705  & M1 & 0.47 & 2.3E+06  & 2.0E-08    \\       
\hline
       \multicolumn{8}{c}{\bf{Chamaeleon II}} \\
\hline
01 & DKCha             &   18.62  & 7200 & F0 & 1.92  & 6.8E+06  & 4.7E-07     \\
02 & IRAS12535-7623    &   1.38  & 3850 & M0 & 0.55  & 7.9E+05  & 3.0E-08     \\
03 & ISO ChaII 28      &   15.85  & 4500 & K4 & 1.88  & 2.9E+05  & 4.0E-07     \\
04 & Sz49              &  0.20  & 3777 & M0 & 0.52   & 7.2E+06    & 4.0E-09     \\
05 & Sz48SW            &  0.26  & 3705 & M1 & 0.46  & 3.9E+06  & 9.6E-09     \\
06 & Sz50              &   1.15  & 3415 & M3 & 0.32   & 2.0E+05    & 8.4E-08     \\
07 & CMCha             &  0.72  & 4060 & K7 & 0.76  & 2.8E+06  & 9.4E-09     \\
08 & IRAS13005-7633    &  0.23  & 3560 & M2 & 0.38  & 3.2E+06  & 6.0E-09     \\
09 & Hn23              &  0.87  & 4350 & K5 & 1.17  & 5.0E+06  & 5.8E-09     \\
10 & Hn24              &   1.05  & 3850 & M0 & 0.56  & 1.0E+06  & 1.9E-08     \\
11 & Sz53              &  0.32  & 3705 & M1 & 0.47  & 3.0E+06  & 7.4E-09     \\
12 & Sz56              &  0.34  & 3270 & M4 & 0.27   & 1.55E+06   & 8.1E-09     \\
13 & Sz57              &  0.41  & 3125 & M5 & 0.22  & 2.9E+05  & 2.4E-08     \\
14 & Sz58              &  0.69  & 4350 & K5 & 1.10    & 7.1E+06    & 5.0E-09     \\
15 & Sz61              &   1.17  & 4350 & K5 & 1.14  & 3.1E+06  & 2.5E-08     \\
16 & IRASF13052-7653NW &  0.20  & 3777 & M0 & 0.52   & 7.2E+06    & 4.4E-09     \\
17 & IRASF13052-7653N  &  0.34  & 3632 & M1 & 0.42  & 2.3E+06  & $<$8.0E-09  \\
\hline
       \multicolumn{8}{c}{\bf{L1641}} \\ 
\hline
01 & [CHS2001]13811 &    0.89     & 4350 & K5 & 1.17 & 4.8E+06  & 2.4E-08     \\
02 & [CTF93]50      &   11.55     & 4200 & K6 & 1.10 & 2.4E+05  & 4.2E-07     \\
03 & [CTF93]47      &   40.93     & 9230 & A1 & 2.35  & 4.5E+06    & 7.3E-07     \\
04 & [CTF93]32      &    2.46     & 3850 & M0 & 0.55 & 5.2E+05  & 1.3E-07     \\
05 & [CTF93]72      &    2.02     & 4350 & K5 & 1.10 & 1.4E+06  & 1.0E-07     \\
06 & [CTF93]83      &    3.68     & 4060 & K7 & 0.75 & 5.2E+05  & 1.2E-07     \\
07 & [CTF93]62      &   10.63     & 8200 & A6 & 1.81 & 1.8E+07  & 4.4E-08     \\
08 & [CTF93]79      &  127.29     & 13000& B7 & 3.50   & 4.0E+06    & 1.8E-06     \\
09 & [CTF93]99      &    4.02     & 4590 & K4 & 1.60 & 1.2E+06  & 1.0E-07     \\
10 & [CTF93]87      &    2.16     & 3850 & M0 & 0.56  & 5.5E+05    & 1.4E-07     \\
11 & [CTF93]104     &    3.73     & 5250 & K0 & 1.78 & 5.9E+06  & 1.8E-07     \\
12 & Meag31         &    0.96    & 2800 & M7 & 0.15 & 1.0E+04  & 2.0E-07     \\
13 & [CTF93]146-2   &    2.64    & 3705 & M1 & 0.46 & 1.1E+05  & 1.2E-07     \\
14 & [CTF93]146-1   &    1.40      & 3890 & K7 & 0.58 & 8.1E+05  & 6.4E-07     \\
15 & [CTF93]211     &    2.64    & 4060 & K7 & 0.73 & 6.7E+05  & 1.2E-07     \\
16 & [CTF93]187     &    9.67    & 6280 & F7 & 1.72 & 8.4E+06  & 6.9E-08     \\
17 & [CTF93]168     &    3.94$^b$ & 4350 & K5 & 1.14 & 6.7E+05  & 2.5E-07     \\
18 & [CTF93]191     &    3.78     & 5080 & K1 & 1.90 & 4.2E+06  & 9.2E-08     \\
19 & [CTF93]246A    &    0.21$^b$ & 3850 & M0 & 0.58  & 9.0E+06    & 2.2E-08     \\
20 & [CTF93]186     &    3.19$^b$ & 4350 & K5 & 1.11 & 8.3E+05  & 3.3E-07     \\
21 & [CTF93]246B    &    2.34     & 3850 & M0.& 0.56  & 5.3E+05    & 2.1E-07     \\
22 & [CTF93]237-2   &    0.09     & 3850 & M0 & 0.56& 3.3E+07  & $<$2.4E-09  \\
23 & [CTF93]237-1   &    4.96    & 3850 & M0 & 0.60 & 5.9E+04  & 1.1E-06     \\
24 & [CTF93]216-1   &   12.44$^b$ & 3560 & M2 & 0.47  & $<$1.0E+04 & 1.9E-06     \\
25 & [CTF93]216-2   &    0.83$^b$ & 3200 & M5 & 0.24 & 1.1E+05  & 2.9E-06     \\
26 & [CTF93]245B-2  &    0.90      & 4350 & K5 & 1.17 & 4.7E+06  & 7.0E-08     \\
27 & [CTF93]245B-1  &    1.66     & 4200 & K6 & 0.89 & 1.2E+06  & 4.7E-08     \\
\hline
\end{tabular}
\end{tiny}
\\~
Notes. $^a$ \macc\ computed from \pab. $^b$ $I$ magnitude not available for ST spectrum normalisation, $J$ magnitude used (see appendix).
\end{table*}

\begin{table*}[!t]
\caption{\label{tab:hr2} Same as Table \ref{tab:hr1} for the POISSON sources of Ser and Lup.}
\begin{tiny}
\begin{tabular}{clcccccc}
\hline
\hline
ID  &  Name  &  \lstar(\lsun)  &  \teff(K)  & ST &  \mstar(\msun) &  age(yr) & \macc(\msunyr) \\
\hline
       \multicolumn{8}{c}{\bf{Serpens}} \\ 
\hline
01 & [WMW2007]103 &    1.37 & 4832 & K2 & 1.39 & 8.2E+06    &  9.7E-09                   \\
02 & [WMW2007]65  &   0.21 & 4060 & K7 & 0.77 & 2.1E+07    &  $<$8.3E-10                \\
03 & [WMW2007]7   &    1.90 & 3400 & M3 & 0.32  & 5.0e+04      &  $<$1.9E-08                \\
04 & [WMW2007]81  &   0.48 & 3560 & M2 & 0.38  & 1.5e+06      &  1.8E-08$^a$                  \\
05 & [WMW2007]38  &    1.00 & 3199 & M5 & 0.24 & 7.9E+04    &  4.0E-08                   \\
06 & [WMW2007]80  &   0.49 &  ... & ...& ...   & ...          & ...                              \\
07 & [WMW2007]85  &    2.00 & 2965 & M6 & 0.18  & $<$1.0e+04   &  $<$1.8E-07                \\
08 & [WMW2007]35  &   0.27 & 3270 & M4 & 0.26 & 2.05E+06    &  $<$2.1E-07                \\
09 & [WMW2007]83  &    4.40 & 4900 & K2 & 2.00 & 2.4E+06    &  4.3E-08                   \\
10 & [WMW2007]70  &    1.00 & 3600 & M2 & 0.40   & 8.5e+05      &  $<$3.0E-11                \\
11 & [WMW2007]37  &    4.30 & 4000 & K7 & 0.70 & 5.2E+05    &  3.2E-07                   \\
12 & [WMW2007]2   &     ... & ...  & ...& ...   & ...          & ...                               \\
13 & [WMW2007]27  &    0.96 & 3997 & K7 & 0.69 & 1.7E+06    &  $<$5.1E-08                \\
14 & [WMW2007]4   &     ... & ...  & ...& ...   & ...          & ...                               \\
15 & [WMW2007]10  &    9.50 & 3872 & K7 & 0.68 & 2.1E+04    &  2.0E-07                   \\
16 & [WMW2007]78  &   0.54 & 3850 & M0 & 0.57 & 2.2E+06    &  4.4E-09$^a$                  \\
17 & [WMW2007]73  &   0.78 & 3487 & M2 & 0.35 & 9.5E+05    &  $<$1.9E-08                \\
\hline
       \multicolumn{8}{c}{\bf{Lupus}} \\ 
\hline
01 & Sz65                   &  0.85 & 3800 & M0 & 0.52 & 1.1E+06   &                   $<$9.9E-09        \\
02 & Sz66                   &  0.20 & 3415 & M3 & 0.31 & 2.8E+06   &  2.3E-09                            \\
03 & [MJS2008]14            &  0.70 & 4600 & K4 & 1.15  & 1.3E+07     &  6.3E-09                            \\
04 & Sz68                   &   4.82 & 4955 & K1 & 2.00 & 2.5E+06   &                   $<$1.5E-08        \\
05 & Sz69                   & 0.09 & 3197 & M5 & 0.19 & 4.2E+06   &  2.7E-09                            \\
06 & [MJS2008]17            & 0.09  & 2700 & M7 & 0.06  & 2.5E+05     &                   $<$9.5E-09        \\
07 & Sz71                   &  0.31 & 3632 & M1 & 0.42 & 2.6E+06   &                   2.8E-09$^a$          \\
08 & Sz72                   &  0.25 & 3560 & M2 & 0.38 & 2.9E+06   &  7.1E-09                            \\
09 & Sz73                   &  0.42 & 4060 & K7 & 0.80 & 6.6E+06   &  1.0E-08                            \\
10 & Sz75                   &   1.50 & 3900 & K7 & 0.59 & 7.7E+05   &  3.4E-08                            \\
11 & Sz82                   &   1.29& 3800 & M0 & 0.52 & 7.9E+05   &                   $<$3.3E-08        \\
12 & Sz83                   &   1.31 & 4060 & K7 & 0.74 & 1.3E+06   &  9.0E-08                            \\
13 & Sz84                   &  0.12 & 3125 & M5 & 0.18  & 2.9E+06     &  4.3E-09                            \\
14 & RY Lup                 &   1.26 & 4590 & K4 & 1.35 & 5.5E+06   &                   2.4E-09$^a$          \\
15 & [MJS2008]146           &   2.74 & 2935 & M6 & 0.18  & $<$1.0E+04  &  7.8E-07                            \\
16 & [MJS2008]149           &   1.45 & 2820 & M7 & 0.15  & $<$1.0E+04  & 2.7E-07                                \\
17 & [HHC93] F403           &   1.79 & 5152 & K0 & 1.39 & 1.1E+07   &                 $<$2.2E-09                        \\
18 & EX Lup                 &  0.47 & 3802 & M0 & 0.53 & 2.3E+06   &  7.1E-09                            \\
19 & Sz133                  &  0.36 & 4400 & K4 & 0.93 & 2.1E+07   &                   1.4E-09$^a$          \\
20 & Sz88                   &  0.49 & 3850 & M0 & 0.57 & 2.5E+06   &  3.5E-08                            \\
21 & Sz88B                  &  0.12 & 3197 & M5 & 0.21& 3.3E+06   &                   7.9E-10$^a$          \\
22 & [MJS2008]20            &  0.30 & 4590 & K4 & 0.86& 3.8E+07   &                   $<$7.3E-10        \\
23 & Sz90                   &   1.10 & 3900 & K7 & 0.59 & 1.0E+06   &  6.2E-09                            \\
24 & Sz95                   &  0.26 & 3400 & M3 & 0.31 & 2.1E+06   &                   $<$3.8E-09        \\
25 & [MJS2008]36            &   2.09 & 2935 & M6 & 0.18  & $<$1.0E+04  &                   $<$3.3E-07        \\
26 & Sz96                   &  0.82 & 3560 & M2 & 0.39 & 9.6E+05   &                   7.5E-09$^a$          \\
27 & Sz97                   &  0.16 & 3270 & M4 & 0.25 & 3.0E+06   &                   $<$4.2E-09        \\
28 & Sz98                   &   2.35 & 4350 & K5 & 1.11 & 1.2E+06   &                   5.9E-09$^a$          \\
29 & Sz99                   & 0.07 & 3270 & M4 & 0.21 & 6.0E+06   &                   1.8E-09$^a$          \\
30 & Sz100                  &  0.17 & 3057 & M5 & 0.17 & 7.3E+05   &                   $<$6.7E-09        \\
31 & Sz103                  &  0.18 & 3270 & M4 & 0.25 & 2.7E+06   &                   $<$3.6E-09        \\
32 & [MJS2008]50            &   3.03 & 5000 & K1 & 1.79 & 4.5E+06   &                   $<$6.1E-09        \\
33 & Sz104                  &  0.10 & 3125 & M5 & 0.17 & 3.7E+06   &                   $<$4.0E-09        \\
34 & HR5999                 &   48.2 & 7890 & A6 & 2.55 & 3.4E+06   &  3.8E-07                           \\
35 & Sz106                  &  0.10 & 3777 & M0 & 0.51 & 2.0E+07   &  6.2E-10                            \\
36 & Sz107                  &  0.15 & 2935 & M5 & 0.12 & 4.0E+05   &                   $<$8.2E-09        \\
37 & Sz109                  &  0.14 & 2800 & M7 & 0.09  & 2.5e+05     &                   $<$4.4E-09            \\
38 & [CFB2003]Par-Lup3-3    &  0.23 & 3270 & M4 & 0.26 & 2.3E+06   &                   $<$5.2E-09        \\
39 & Sz110                  &  0.28 & 3270 & M4 & 0.27  & 1.8e+06     &  4.8E-09                                \\
40 & [MJS2008]68            &  0.44 & 3900 & K7 & 0.62 & 3.5E+06   &  1.6E-09                            \\
41 & Sz111                  &  0.33 & 3750 & M0 & 0.50 & 3.3E+06   &  2.9E-09                            \\
42 & Sz112                  &  0.19 & 3125 & M5 & 0.20 & 2.3E+06   &                   $<$9.2E-09        \\
43 & Sz113                  & 0.06 & 3197 & M5 & 0.17 & 5.8E+06   &  3.9E-09                            \\
44 & Sz114                  &  0.31 & 3175 & M5 & 0.23  & 1.55e+06    &  1.7E-08                            \\
45 & Sz117                  &  0.47 & 3700 & M1 & 0.46 & 1.8E+06   &                   $<$5.6E-09        \\
46 & Sz118                  &  0.92 & 4060 & K7 & 0.75 & 1.2E+06   &  6.2E-09                            \\
47 & [MJS2008]113           & 0.05 & 2900 & M6 & 0.09  & 4.2E+06     &                   $<$3.1E-09        \\
48 & [MJS2008]114           & 0.03 & 2990 & M6 & 0.09  & 6.0E+06     &                   $<$1.5E-09        \\
49 & Sz123                  &  0.20 & 3705 & M1 & 0.46  & 5.7E+06     &  7.0E-09                            \\
50 & [MJS2008]133           &   1.00 & 4400 & K4 & 1.20 & 4.7E+06   &                   $<$1.2E-09        \\
51 & [MJS2008]136           &   2.87 & 5300 & G9 & 1.57 & 8.4E+06   &                   9.6E-09$^a$          \\
52 & [MJS2008]137           &  0.60 & 4400 & K4 & 1.10 & 1.0E+07   &                   $<$1.8E-09        \\
\hline
\end{tabular}
\end{tiny}
~\\
Notes. $^a$ \macc\ computed from \pab.
\end{table*}

\section*{Appendix C: example spectra.}
\label{apx:spectra}

\begin{figure*}[] 
\centering
\includegraphics[width=19cm]{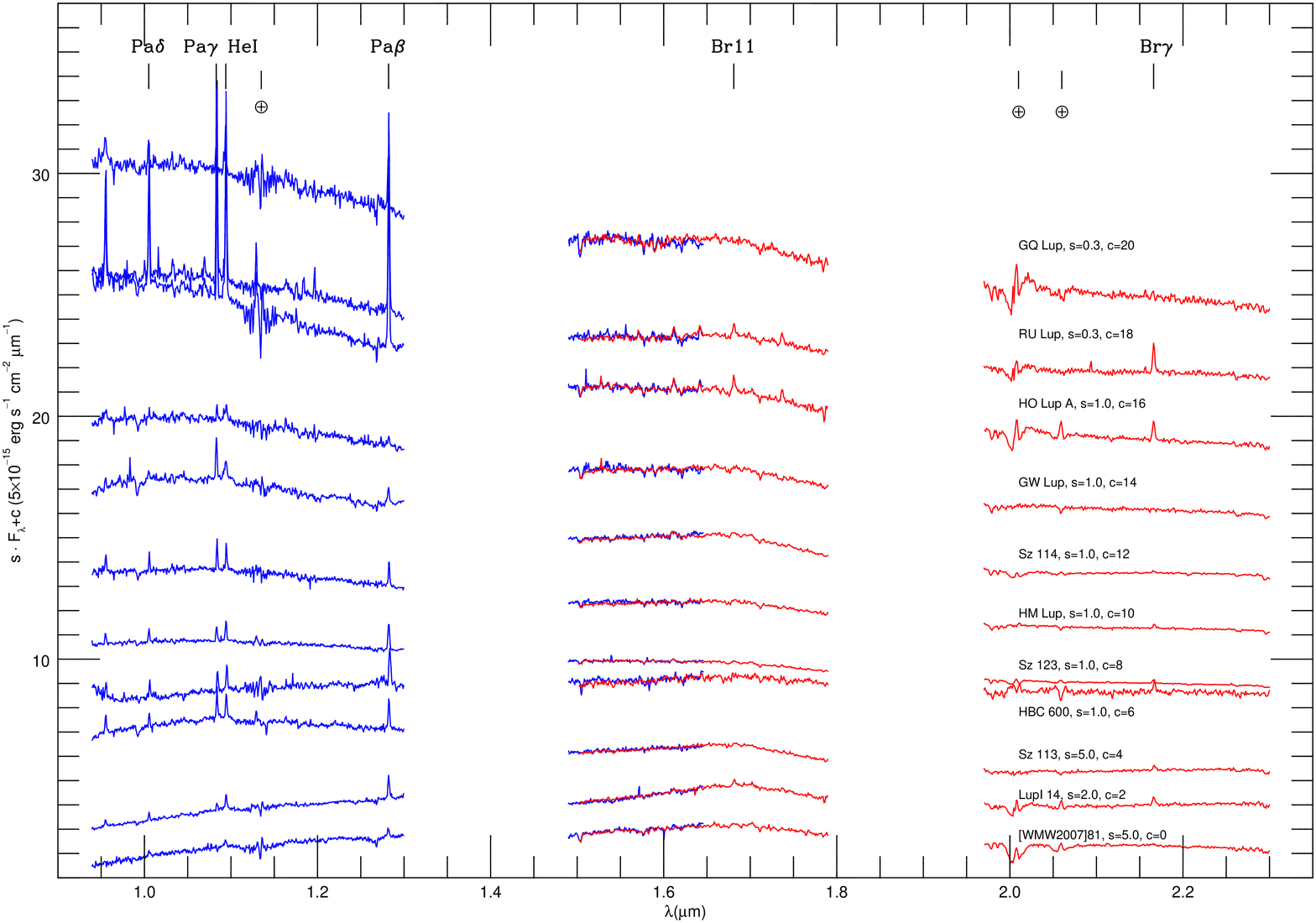}\\
\caption{\label{fig:spectra} SofI near-IR spectra for a few selected Lupus and Serpens sources. Different colours refer to the two SofI grisms (blue and red). Spectra are shown in flux units of $5\times10^{-15}$ erg s$^{-1}$ cm$^{-2}$ $\mu$m$^{-1}$ and were offset by a constant $c$ and multiplied by a scale factor $s$ (both indicated) for better visualisation. Wavelength intervals heavily corrupted by atmospheric absorption were removed. The position of the main emission lines present in the covered spectral range is indicated, as well as residuals from atmospheric features.
}
\end{figure*}

A few representative POISSON spectra of Lupus and Serpens sources are displayed in Fig.~\ref{fig:spectra}, showing sources with different line-to-continuum ratios.

%
\end{document}